\documentclass[sigconf]{acmart}

\usepackage{fancyhdr}
\usepackage{flushend}

\usepackage{xspace}
\usepackage{enumitem}
\usepackage{soul}
\usepackage{def}
\usepackage[ruled,linesnumbered]{algorithm2e} 
\usepackage{multirow}
\usepackage{makecell}
\usepackage{graphicx}
\usepackage{textcomp}
\usepackage{tikz}

\usepackage{listings}
\usepackage{cleveref}

\AtBeginDocument{%
  }

\copyrightyear{2023}
\acmYear{2023}
\setcopyright{acmlicensed}\acmConference[MICRO '23]{56th Annual IEEE/ACM International Symposium on Microarchitecture}{October 28-November 1, 2023}{Toronto, ON, Canada}
\acmBooktitle{56th Annual IEEE/ACM International Symposium on Microarchitecture (MICRO '23), October 28-November 1, 2023, Toronto, ON, Canada}
\acmPrice{15.00}
\acmDOI{10.1145/3613424.3614309}
\acmISBN{979-8-4007-0329-4/23/10}

\begin{document}

\title{G10: Enabling An Efficient Unified GPU Memory and Storage Architecture with Smart Tensor Migrations}


%
%

\author{Haoyang Zhang}
\authornote{Co-primary authors.}
\email{zhang402@illinois.edu}
\affiliation{%
  \institution{UIUC}
  \city{}
  \state{}
  \country{}
}

\author{Yirui Eric Zhou}
\authornotemark[1]
\email{yiruiz2@illinois.edu}
\affiliation{%
  \institution{UIUC}
  \city{}
  \state{}
  \country{}
}

\author{Yuqi Xue}
\email{yuqixue2@illinois.edu}
\affiliation{%
  \institution{UIUC}
  \city{}
  \state{}
  \country{}
}

\author{Yiqi Liu}
\email{yiqiliu2@illinois.edu}
\affiliation{%
  \institution{UIUC}
  \city{}
  \state{}
  \country{}
}

\author{Jian Huang}
\email{jianh@illinois.edu}
\affiliation{%
  \institution{UIUC}
  \city{}
  \state{}
  \country{}
}

\renewcommand{\shortauthors}{Haoyang Zhang, Yirui Eric Zhou, Yuqi Xue, Yiqi Liu, and Jian Huang}

\begin{abstract}
To break the GPU memory wall for scaling deep learning workloads, a variety of architecture and system techniques have 
been proposed recently. Their typical approaches include memory extension with flash memory and direct storage access. 
However, these techniques still suffer from suboptimal performance and introduce 
complexity to the GPU memory management, making them hard to meet the scalability requirement of deep 
learning workloads today. 

In this paper, we present a unified GPU memory and storage architecture named \pname{} driven by the fact that 
the tensor behaviors of deep learning workloads are highly predictable. \pname{} integrates the host memory, 
GPU memory, and flash memory into a unified memory space, to scale the GPU memory capacity while 
enabling transparent data migrations. Based on this unified GPU memory and storage architecture, 
\pname{} utilizes compiler techniques to characterize the tensor behaviors in deep learning workloads. 
Therefore, it can schedule data migrations in advance by considering the available bandwidth of flash memory and host memory. 
The cooperative mechanism between deep learning 
compilers and the unified memory architecture enables \pname{} to hide data transfer overheads in a 
transparent manner. We implement \pname{} based on an open-source GPU simulator. Our experiments 
demonstrate that \pname{} outperforms state-of-the-art GPU memory solutions by up to 1.75$\times$, without code modifications 
to deep learning workloads. With the smart data migration mechanism, \pname{} can reach
90.3\% of the performance of the ideal case assuming unlimited GPU memory.

\end{abstract}




\maketitle

\section{Introduction}
\label{sec:intro}

As we utilize GPUs for scaling deep learning workloads with large-scale data sets,
we are facing the well-known memory wall~\cite{zng:isca2020, tensordimm, BaM}. 
Although GPUs provide increasing parallelism, their on-board 
memory capacity is still limited, due to the space and power constraints, as well as DRAM scaling 
issues~\cite{tensordimm, memorywall:micro2018, neummu:asplos2020, jain:mlsys2020}.
Meanwhile, the deep neural network (DNN) models, which have become the killer 
applications of GPUs, are demanding a growing amount of memory for training efficiency and scalability~\cite{devlin2018bert, tan2019efficientnet, 
szegedy2017inception, zagoruyko2016wide, SENet, ResNeXt, szegedy2016rethinking, ResNet,NEURIPS2020_1457c0d6}. 
This gap will only be enlarged if not addressed properly.


To overcome the GPU memory wall, a promising and practical approach is to expand the limited GPU memory with flash memory, 
which provides larger memory capacity at a low cost~\cite{zng:isca2020, zero_infinity}.
With this approach, a few architecture solutions have been developed in both academic~\cite{zng:isca2020, flashgpu} and industry~\cite{ssg}.
For example, ZnG directly replaces the GPU on-board DRAM with low-latency flash chips~\cite{zng:isca2020}, and AMD SSG 
integrates flash-based solid-state drives (SSDs) into the GPU board~\cite{ssg}.
Unfortunately, the limited bandwidth of flash chips is still the performance bottleneck, in comparison with the high-bandwidth 
memory in GPUs~\cite{h100-hbm}.
An alternative approach is to use off-board flash-based SSD to back the GPU on-board memory, forming a 
heterogeneous memory and storage system. 
For example, GPU vendors have been connecting GPUs with SSDs 
via PCIe links to bypass the host CPU, and allowing direct data transfer between the SSD and GPU~\cite{ssg,ssg-detail,gpudirectstorage}.

However, these existing solutions still suffer from suboptimal performance. 
Although we can scale up the SSD bandwidth by stacking 
multiple SSDs or flash chips, the aggregated bandwidth is still limited by the PCIe interface. 
Even though we can employ faster interconnects such as NVLink~\cite{ang:gpulink}, the data transfer bandwidth
is still much lower than the GPU on-board memory bandwidth. 
To tolerate slow flash accesses, developers have to carefully manage the data across the heterogeneous 
memories to explore the data locality~\cite{flashneuron, SwapAdvisor, deepum, autotm}. 
This inevitably complicates the GPU memory management and hurts the development productivity. 

Ideally, we wish to transparently expand the GPU memory using low-cost flash memory, while achieving similar performance as that of the GPU with unlimited on-board DRAM.
Our characterization study of diverse DNN models (see $\S$\ref{sec:study}) shows that this is feasible. 
We disclose that (1) only a small portion (less than 10\%) of tensors are active in each DNN training iteration, and (2) a majority of inactive tensors remain inactive for a long period of time (see Figure~\ref{fig:cold_period_length_cdf}).
This offers sufficient opportunities for us to move tensor data across heterogeneous memory devices. Therefore, if we can intelligently 
move inactive tensors from the fast GPU memory to the slow memories (i.e., host memory and flash memory), 
we can not only improve the utilization of the precious GPU memory but also hide the data access overheads of the slow memories. 

To achieve the aforementioned goals, we have to overcome three major challenges. First, to enable intelligent 
tensor migrations, we need to capture the memory demand and lifetime of different tensors in a deep learning model. 
The tensor-level semantic knowledge will serve as the guidance for scheduling tensor migrations. Second, as different 
tensors have different properties (i.e., tensor size and lifetime in Figure~\ref{fig:inactive_distriubtion}), 
we need to carefully decide which tensor should be migrated, where it should be migrated to, and when it should be migrated. 
Third, the tensor migrations should be transparent to applications, and the migration should be executed in an automated manner without requiring manual effort from developers. 

In this paper, we present \pname{}, a unified GPU memory and storage architecture that enables smart tensor migrations for 
scaling the GPU memory transparently using flash memory, while tolerating the performance overheads of slow flash accesses.  
\pname{} consists of three major components: (1) a tensor vitality analyzer for extracting the semantic knowledge 
of tensors in a deep learning model, (2) a tensor migration scheduler for planning the tensor migrations 
in advance, and (3) a unified memory system for simplifying the GPU memory management and enabling transparent tensor migrations.  

The tensor vitality analyzer works with deep learning frameworks like PyTorch to track all the tensors in a DNN model. 
It leverages the execution graph generated by the compiler to learn the size and lifetime of each tensor as well as its 
dependency on other tensors. Therefore, the analysis procedure is almost free at the compilation stage. 
Based on the extracted semantic knowledge of tensors, 
the tensor migration scheduler of \pname{} will plan the tensor migrations in advance before executing the model training process. 

To maximize the benefits of tensor migrations, \pname{} prefers to migrate large tensors that will be inactive for a long time to 
the flash memory. Therefore, the precious GPU memory can be best utilized for active tensors. 
\pname{} will migrate these inactive tensors as many as possible to fully utilize the available bandwidths of flash memory and host memory.
For the inactive tensors whose inactive time is short, \pname{} will make the best effort to keep them in the GPU memory 
to avoid unnecessary tensor migrations. In order to tolerate the long access delay of flash memory and host memory, 
\pname{} also plans intelligent data prefetching in advance with its tensor migration scheduler.
The detailed algorithms of the tensor migration scheduler will be discussed in $\S$\ref{sec:design}. 

To facilitate the tensor migration, \pname{} integrates the GPU memory, host memory, and flash memory as a unified memory space by 
extending the Unified Virtual Memory (UVM)~\cite{analysis-UVM} of GPUs. G10 extends the page table of UVM by storing flash page addresses in its leaf-level page table entries. The unified page table can point to an address in either host memory, GPU memory, or flash memory. As \pname{} plans tensor migrations, it only needs to specify the virtual addresses of tensors.
The unified memory system will conduct the transparent address translation at runtime. This significantly simplifies the GPU memory management and the compiler optimizations.  

We implement \pname{} by extending an open-source GPU simulator UVMSmart~\cite{InterplayUVM}.
To evaluate \pname{}, we run a variety of DNN models with different batch sizes.
Compared to state-of-the-art solutions,
\pname{} improves the end-to-end DNN training performance by up to 1.75$\times$,
while scaling the GPU memory with low-cost flash memory.
With smart tensor migrations planned at the compilation stage,
\pname{} delivers 90.3\% of the performance of the ideal case assuming unlimited GPU memory.
Our sensitivity analysis shows that \pname{} still has significant benefits, 
as we scale the GPU-SSD PCIe bandwidth.
Overall, we make the following contributions:

\begin{itemize}[itemsep=0pt,topsep=0pt,parsep=0.2em,partopsep=0.2em,leftmargin=*]
\vspace{1ex}
\item We conduct a characterization study of the memory usage of diverse DNN training workloads, and show that the predictable tensor behaviors of DNN models provide sufficient opportunities for enabling smart tensor migrations. 
\vspace{0.5ex}
\item We develop a unified GPU memory and storage architecture named \pname{}, and show the feasibility of scaling GPU memory with flash memory, while achieving similar performance as the ideal case assuming unlimited GPU memory. 
\vspace{0.5ex}
\item We propose a smart tensor migration mechanism that can intelligently plan tensor migrations across heterogeneous memories at the compilation stage, based on the extracted semantic knowledge of tensors. 
\vspace{0.5ex}
\item We evaluate \pname{} against state-of-the-art GPU memory solutions and show its benefits for various DNN models. 
\vspace{1ex}
\end{itemize}

\section{Background and Motivation}
\label{sec:background}
In this section, we first present modern GPU memory and storage architecture. After that, we 
discuss existing approaches to scaling GPU memory, and their limitations.  

\begin{figure}[t]
    \centering
    \includegraphics[width=0.8\columnwidth]{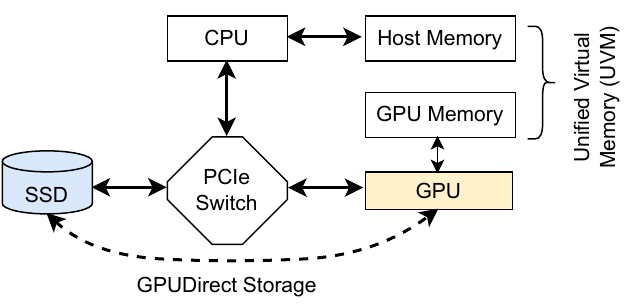}
    \caption{Modern GPU memory/storage architecture.}
    \label{fig:gpuarch}
\vspace{-1ex}
\end{figure}

\subsection{GPU Memory and Storage Architecture}
\label{subsec:gpuarch}

We demonstrate the system architecture of modern GPU memory and storage in Figure~\ref{fig:gpuarch}.
The GPU and storage devices like SSDs are connected with the host machine through the
Peripheral Component Interconnect Express (PCIe)~\cite{pcisig}. 
While GPU has its on-board memory, the memory capacity is constrained by the DRAM scaling wall and
the limited on-board space for memory packages~\cite{zng:isca2020}. Therefore, their memory cannot host
the entire working set of large-scale deep learning workloads. To address this problem, GPUs 
follow the same way of managing memory/storage devices in CPU-centric computing, and use the storage device 
as a swapping disk. If a page requested by the GPU is not in its memory, a page fault will happen. 
And the GPU will inform the host to handle the page fault, load the page from the storage device, 
and move it to the GPU memory, causing significant data movement overhead.



%
%
%

\subsection{Approaches to Scaling GPU Memory}
\label{subsec:existingapproach}


\vspace{0.2em}
\noindent
\textbf{Expand GPU memory with host memory.}
Compared to the GPU memory, the host machine usually equips a larger memory with limited bandwidth, 
making it a natural option for expanding GPU memory. 
While developers can manually swap the data between the host and GPU, modern GPUs make this procedure transparent 
with unified virtual memory (UVM)~\cite{uvm, amduvm}. UVM enables a unified and coherent virtual memory space between the host and GPU, so application data can be allocated to the space and accessed by host and GPU with shared virtual addresses.
With the cooperation of GPU hardware and runtime, UVM maintains data consistency transparently and enables on-demand data migrations between the host and GPU at page granularity.

Upon accessing a UVM page absent in the GPU memory, a GPU page fault
will be triggered to request a data migration from the host~\cite{towardsUVM,analysis-UVM}.
When the GPU memory is fully occupied, the least recently used pages are evicted from the GPU memory to the host memory. 
To improve the swapping efficiency, prior studies~\cite{SwapAdvisor,InterplayUVM,sentinel,autotm,le2019tflms,deepum,zero_infinity,oc-dnn} 
developed optimization techniques for improved data locality. However, GPU memory still cannot scale purely relying on the host memory 
to meet the increasing demands of deep learning workloads, especially those large ones. 


\vspace{0.2em}
\noindent
\textbf{Expand GPU memory with flash memory.}
An alternative approach is to expand GPU memory with SSDs, as shown in Figure~\ref{fig:gpuarch}. 
The rapidly shrinking process technology has allowed SSDs to boost their bandwidth and capacity by increasing the number
of chips. However, the GPU has to communicate with the host CPU to access data on the SSD, which incurs significant 
performance overhead~\cite{gpufs, gpunet, activepointer:isca2016}.
Most recently, NVIDIA's GPUDirect Storage allows GPU
to bypass the host CPU and directly access the SSD via the PCIe interface~\cite{gpudirectstorage}.
AMD's DirectGMA~\cite{directgma} also enables a similar functionality.

However, current approaches of using flash memory 
to expand GPU memory are still suffering from suboptimal performance, as they cannot efficiently hide the 
slow flash accesses. A recent study proposed to offload intermediate data of DNN models to the SSD~\cite{flashneuron}, and 
overlap the GPU processing with flash data accesses. However, due to the lack of rich semantic knowledge of 
tensors, there is still much space for improvement. In this paper, we conduct a characterization study of the semantic knowledge of 
tensors, and demonstrate the unexplored opportunities in $\S$\ref{sec:study}. 

\section{GPU Memory Characterizations}
\label{sec:study}

\begin{figure}[t]
    \centering
    \includegraphics[width=0.9\columnwidth]{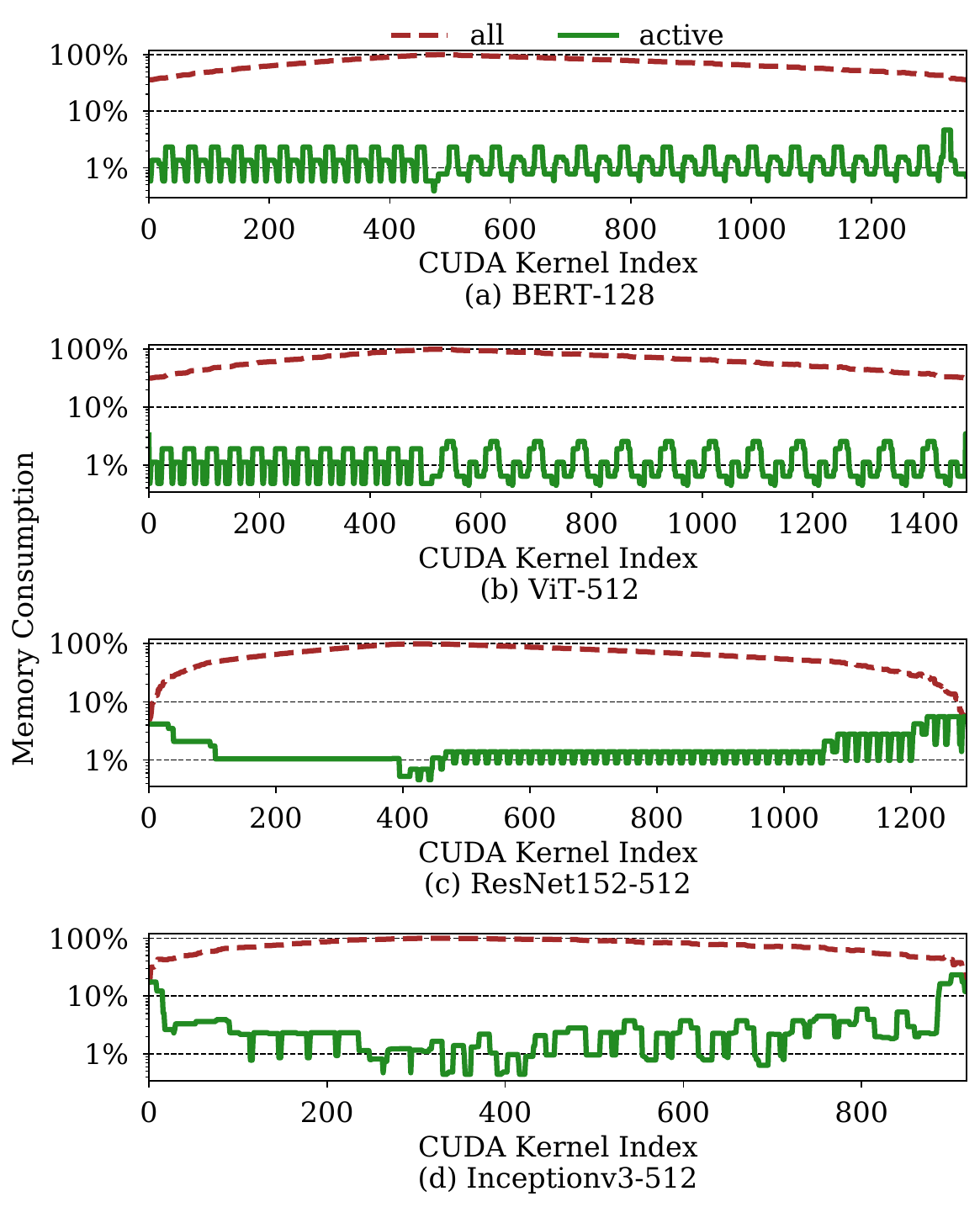}
	\caption{Memory consumption of all and active tensors (w.r.t. peak memory consumption in a single training iteration). CUDA kernel indexes are in execution order.}
    \label{fig:proportion_tensor}
\end{figure}

In this section, we first study the memory usage patterns of DNN training for representative real-world large models listed in Table~\ref{table:benchmarks}. We analyze the DNN dataflow graph to extract useful DNN semantics and profile the execution of each CUDA kernel on an NVIDIA A100 GPU.
For ease of discussion, we define that a tensor is \textit{active} at a certain time if it is used by the currently executing kernel, or \textit{inactive} otherwise.
We summarize our findings as follows.


\vspace{0.2em}
\noindent
\textbf{Small memory requirement of active tensors.}
We first study the total memory demand of a single training iteration. 
Figure~\ref{fig:proportion_tensor} shows the amount of GPU memory required by active tensors and the total memory required during a training iteration.
For most DNN models, active tensors only account for less than 10\% (1\% on average) of the total memory requirement.
While the memory capacity required by the entire DNN can greatly exceed GPU memory, 
each layer only accounts for a small portion. For example, the largest kernel in our studied models 
occupies 5.7GB of memory, much smaller than the 40GB available memory of A100. This gives abundant opportunities to leverage the unused memory 
for preparing the tensors required by the next kernel, enabling efficient overlapping of GPU compute and memory swapping. 

\observation{1}{
During DNN training, only a small portion of tensors are active and required in GPU DRAM. Most tensors are inactive and can be swapped out.
}

\begin{figure}[t]
    \centering
    \includegraphics[width=\columnwidth]{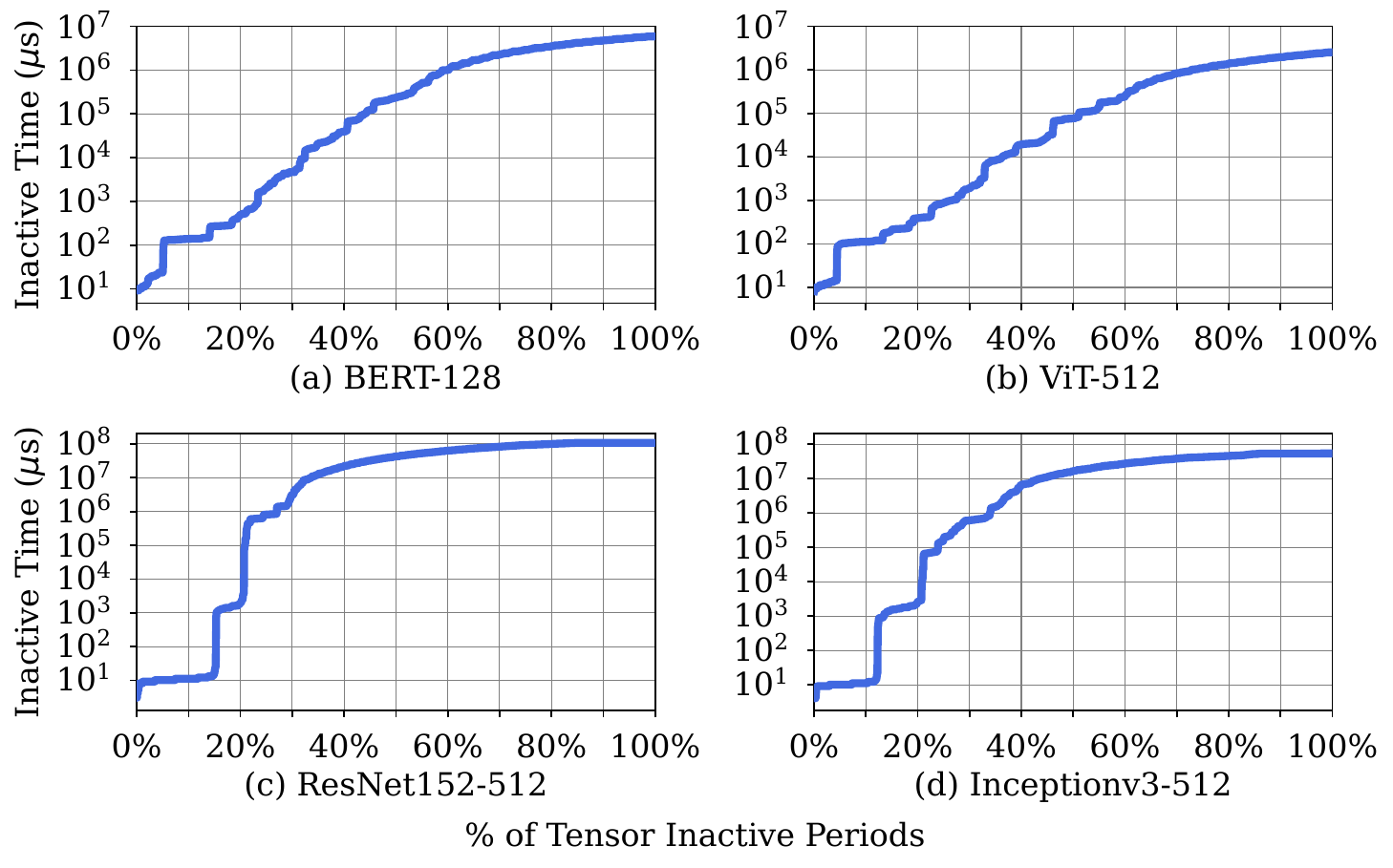}
    \caption{Distribution of tensor inactive period lengths.}
    \label{fig:cold_period_length_cdf}
\end{figure}


\begin{figure}[t]
    \centering
    \includegraphics[width=0.9\columnwidth]{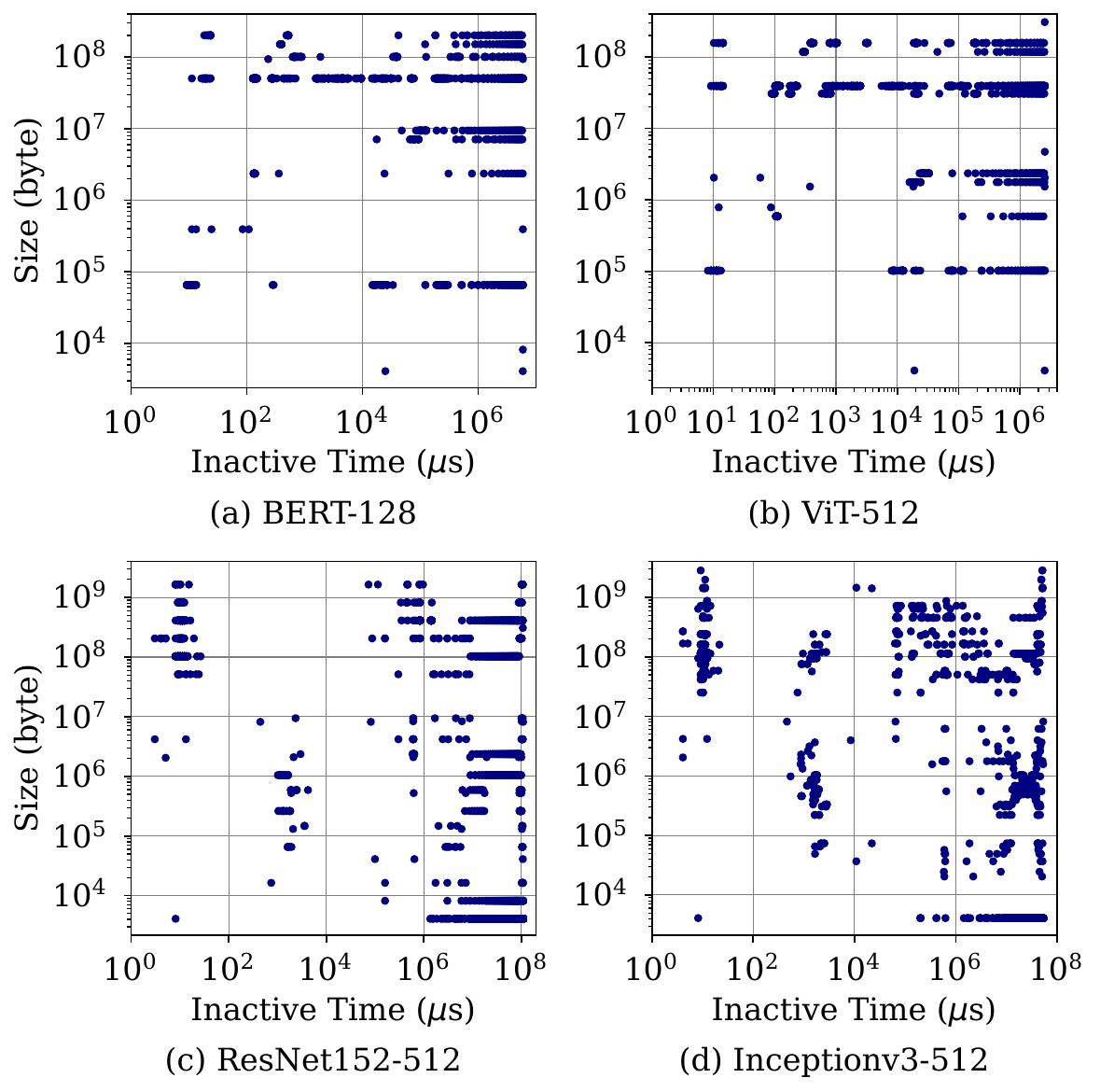}
    \caption{The distribution of inactive periods of tensors having different sizes.
    }
    \label{fig:inactive_distriubtion}
\end{figure}

\noindent
\textbf{Long unused time of inactive tensors.}
To understand the memory usage pattern of inactive tensors, we study how long a tensor remains inactive.
We define an \textit{inactive period} as a time interval during which the tensor remains inactive until it is used by another kernel.
Figure~\ref{fig:cold_period_length_cdf} shows the distribution of lengths of the inactive periods for all tensors. For CNN models (ResNet152 and Inceptionv3), more than 60\% of the inactive periods last longer than $10^7$\textmu s. For Transformer models (BERT and ViT), about 50\% of the inactive periods last longer than $10^5$\textmu s. 
This indicates that many tensors have inactive periods longer than the SSD latency (e.g., 20\textmu s), which provides opportunities for us to swap 
out these tensors to external SSD devices with negligible performance penalties. 

The long unused time of inactive tensors is the result of the temporally sparse tensor access pattern during DNN training. 
In a typical DNN dataflow graph, one tensor only needs to be used twice, one in the forward pass and the other in the backward pass, 
unless the tensor is involved in a branch or join layer. Although the dataflow graphs of some DNN models may have a complex topology consisting of multiple branches, joins, and unrolled loops, the overall dataflow still tends to be linear, so each tensor is only used for a few times.

\observation{2}{
During DNN training, many tensors stay inactive for a long time period. They can be safely swapped out before being needed again by any kernel.
}

\noindent
\textbf{Diversity of inactive tensors.}
Figure~\ref{fig:inactive_distriubtion} shows that the inactive periods of tensors have diverse lengths (e.g., ranging from $\thicksim$10\textmu s to 100s in Inceptionv3-512). 
The inactive tensors also have vastly different sizes (e.g., from less than 10KB to more than 2.7GB in Inceptionv3-512), 
and their distribution is quite sparse. 
In fact, over 60\% to 80\% of inactive periods 
are able to hide the swapping latency, indicating that we have sufficient opportunities to swap tensors. 

When we decide to swap out a tensor, we can reduce the GPU memory consumption during the tensor's inactive period. 
The diversity of inactive tensors introduces challenges to the swapping algorithm design, as different swapping decisions 
can have different benefits and I/O costs. To maximize the efficiency of memory swapping, it is important to choose those tensors that can reduce the memory usage by the largest amount, for the longest time, and with the lowest I/O cost.


\observation{3}{
Different swapping decisions impact GPU memory consumption differently in both time and space, 
given the different sizes and inactive period lengths of tensors. 
To maximize memory efficiency, we should swap out the most beneficial tensors.
}


\noindent
\textbf{Complexity of scheduling tensor swapping.}
In Figure~\ref{fig:proportion_tensor}, we observe that the memory consumption of a DNN program is not uniform throughout its entire execution. 
As we make tensor swapping decisions, 
the GPU memory consumption pattern also changes as tensors are swapped in or out at runtime. 
Moreover, each swap occupies bandwidth of GPU-Host and GPU-SSD communications.
Consequently, the above complexities render a static policy ineffective for deciding which tensor should be evicted and what time this eviction should occur. 

\observation{4}{
The GPU memory consumption changes throughout the DNN training process and is affected dynamically by tensor swapping decisions.
Hence, a static tensor swapping policy is insufficient for finding a globally optimized swapping plan.
}
\vspace{-2ex}
 
\section{\pname{} Design}
\label{sec:design}

\begin{figure*}[!t]
    \centering
    \includegraphics[width=\linewidth]{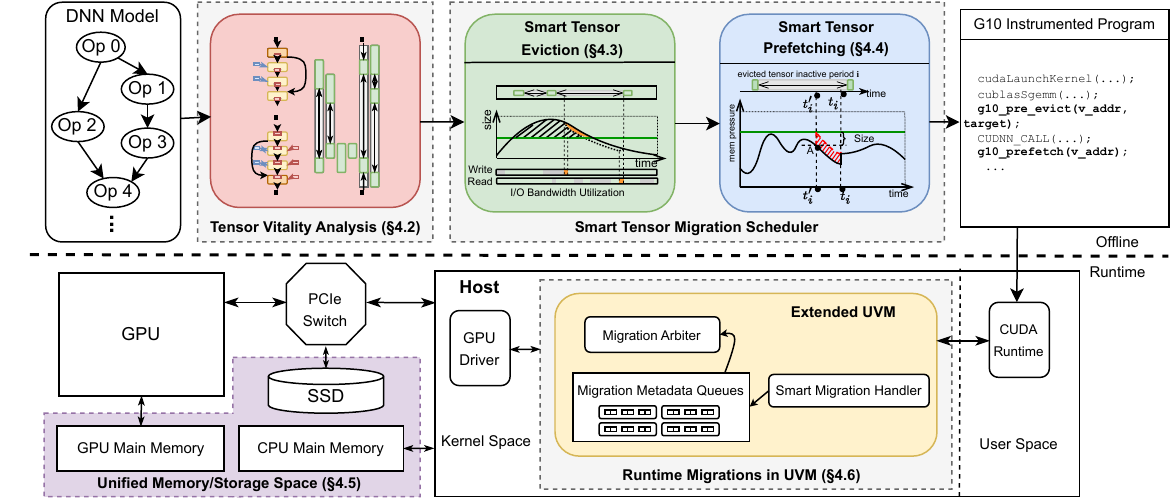}
    \caption{System architecture of \pname{}.}
    \label{fig:overview}
    
\end{figure*}

\subsection{System Overview}
\label{subsec:designoverview}

We show the \pname{} architecture in Figure~\ref{fig:overview}. It has three major components: 
(1) a tensor vitality analyzer that quantifies the tensor size and liveness as we compile a DNN model ($\S$\ref{subsec:tensorAnalysis}); 
(2) a tensor migration scheduler for planning the tensor migrations in advance ($\S$\ref{subsec:tensormigration} and $\S$\ref{subsec:tensorprefetch}); 
and (3) a unified memory system for simplifying GPU memory management and enabling transparent tensor 
migrations ($\S$\ref{subsec:unifiedmemory} and $\S$\ref{subsec:runtime}).  

Given a DNN model, \pname{}'s tensor vitality analyzer will work with DNN compilers to track all the tensors and their 
dependencies, and quantify their sizes and lifetime (i.e., semantic knowledge of tensors). 
With these knowledge, the tensor migration scheduler will 
plan the optimized execution schemes of tensors, with
the goal of maximally overlapping the GPU computation and tensor migrations. 
Identifying a globally optimized tensor migration plan is a dynamic optimization problem, as each tensor migration decision 
will affect subsequent decisions, due to its impact on the GPU memory pressure, and GPU-SSD and GPU-Host bandwidth utilizations. 
Therefore, we used a dynamic algorithm to iteratively find the best tensor candidates for eviction and prefetching.
After that, \pname{} adds the eviction and prefetch instructions into the compiled program. 
GPU will execute these instructions at runtime with the unified GPU memory and storage architecture. 
As the GPU memory, host memory, and SSD are combined into a unified space, the tensor migrations 
are fully transparent to developers and DNN workloads. We will describe each 
component of \pname{} as follows.

\subsection{Tensor Vitality Analysis}
\label{subsec:tensorAnalysis}

\begin{figure*}[t]
    \centering
    \includegraphics[width=\linewidth]{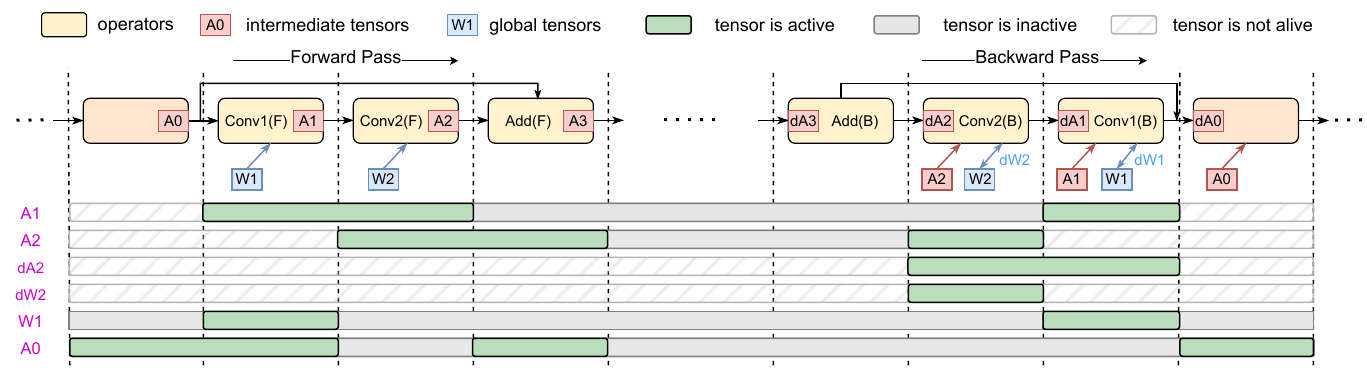}
    \caption{An example of tensor vitality analysis for a residual basic block. Operators in forward propagation and in backward propagation are marked as \texttt{Op(F)} and \texttt{Op(B)}, respectively. \texttt{Ax} and \texttt{Wx} are activation tensors and weight tensors, respectively. \texttt{dAx} and \texttt{dWx} are the corresponding gradient tensors.}
    \label{fig:liveness}

\end{figure*}






\vspace{0.2em}
\noindent
\textbf{Identifying \textit{global tensors} and \textit{intermediate tensors}.}
We first categorize the tensors based on their lifetimes in a DNN training iteration (i.e., one round of forward and backward propagation).
As shown in Figure~\ref{fig:liveness}, a \textit{global tensor} such as model weights (e.g., \texttt{W1}) is used across multiple training iterations. It will be allocated 
in the unified memory space at 
the beginning of the DNN program. 
An \textit{intermediate tensor}, such as the activation and gradient (e.g., \texttt{A1} and \texttt{dA2}), is used within one iteration. We define the tensor as \textit{born} 
the first time when it was used, and as \textit{dead} after the last time it was used. Intermediate tensors can be deallocated after their deaths to free up GPU memory. 

\vspace{0.2em}
\noindent
\textbf{Identifying \textit{tensor inactive time periods}.}
When an operator is being executed on GPU, both its input and output tensors are \textit{active}, and should be present in GPU memory. Otherwise, a tensor is \textit{inactive}, if it is not being used by the currently executing kernel and is not yet dead. We define an \textit{inactive time period} of a tensor as the period during which the tensor is inactive and not dead (i.e., it is not being used right now but will be used in the future). For a complex DNN program, a tensor may have multiple inactive time periods and can be swapped in and out multiple times (e.g., \texttt{W1} and \texttt{A0}).
Both global and intermediate tensors can be inactive, and the inactive time period of a global tensor may span across two consecutive training iterations.
For example, \texttt{W1} turns inactive during the backward pass of the current iteration, and it becomes active again in the forward pass of the next.

The inactive time periods of all tensors indicate when a tensor is safe to be migrated out and when it must be migrated back. As DNN programs have predictable performance and dataflow patterns, \pname{} performs offline compile-time profiling, and uses the execution times of the GPU kernels to estimate the lengths of the inactive time periods. Using the tensor sizes, the storage bandwidth, and the GPU-Host bandwidth, 
\pname{} estimates the eviction and prefetch overheads of each tensor. \pname{} then leverages all the inactive time periods to generate a globally optimized execution plan. 


\subsection{Smart Tensor Eviction}
\label{subsec:tensormigration}


\newcommand\mycommfont[1]{\footnotesize\ttfamily\textcolor{purple}{#1}}
\SetCommentSty{mycommfont}
\SetAlFnt{\small}

\begin{algorithm}[t]
    \DontPrintSemicolon 
    \SetAlgoNoEnd
    \SetAlgoLined
    \KwIn{$gpu\_cap \leftarrow$ the GPU on-board memory capacity \\ 
    \ \ \ \ \ \ \ \ \ \ \ \ $tensors \leftarrow$ 
    the list of all intermediate tensors \\
     \ \ \ \ \ \ \ \ \ \ \ \ $periods \leftarrow$ 
    the list of all tensor inactive periods \\
    }
    \KwOut{A list of \pname{} tensor migration instructions}
    \SetKwFunction{FMain}{$Eviction Scheduling$}
    \SetKwProg{Fn}{Function}{:}{}
    \Fn{\FMain{$gpu\_cap,\  tensors, \ periods$}}{
        {\For{i = 0; i $<$ periods.size; i++}{
            {\If{$max(mem\_pressure) < gpu\_cap$}{
                \textbf{break}
            }}
            sort $periods$ by critical\_mem\_pressure\_reduction) \\
            {\If{$periods[0].critical\_mem\_pressure\_reduction > 0$}{
                $t\_r \leftarrow$ periods[0].start\_time \\
                $t\_s \leftarrow$ periods[0].tensor\_size / BW\_{SSD} \\
                {\If{(to\_ssd\_traffic is full during $t\_r$ to $t\_r$ + $t\_s$)}{
                    {\If{host mem isn't full during periods[0]}{
                        schedule pre-eviction(periods[0].tensor, host) at $t\_r$ \\
                        periods.erase(0) \\
                        update memory pressure and I/O traffic \\  
                        \textbf{continue} \\   
                    }}
                }}
                schedule pre-eviction(periods[0].tensor, SSD) at $t\_r$ \\
                update memory pressure and I/O traffic \\
                periods.erase(0) \\
            }}
        }}
        
    }
    \caption{{Smart Tensor Eviction Algorithm.}} 
    \label{alg:eviction}
\end{algorithm}
\setlength{\textfloatsep}{2pt}

To generate a globally optimized migration plan, the smart tensor eviction algorithm must address the following challenges.
First, we must utilize the limited GPU on-board memory to store the most beneficial tensors. As tensors have different sizes and inactive period lengths, they contribute different degrees of GPU memory pressure. Thus, 
evicting some tensors (e.g., large tensors with long inactive periods) yields more benefits in reducing GPU memory pressure.
Second, we must consider both SSD and host memory as potential migration destinations, as they provide different bandwidths, capacities, and different migration overheads. Ideally, we aim to exploit both the high migration bandwidth of host memory and the large capacity of SSD.
Third, we should best utilize the available migration bandwidth, as DNN workloads are mostly bandwidth-sensitive.
The algorithm should also choose the best timings for tensor migrations. 

To this end, we propose a smart eviction scheduling algorithm that iteratively finds the best eviction candidates (i.e., tensor inactive periods) 
in each training iteration at compile time. The algorithm tracks the GPU memory consumption and the migration bandwidth 
utilization to evaluate potential benefits of an eviction.
We describe its key ideas as follows.



\begin{figure}[!t]
    \centering
    \vspace{-5ex}
    \includegraphics[width=1.02\columnwidth]{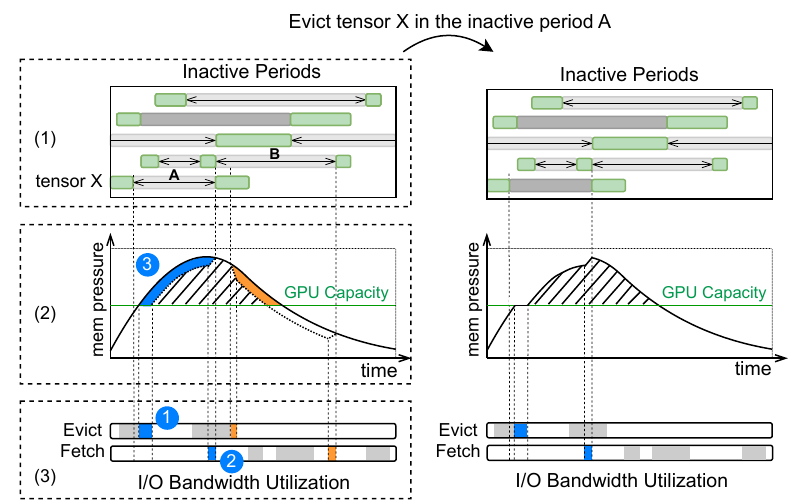}
    \vspace{-4ex}
	\caption{An example of state transition in \pname{}'s smart migration scheduling algorithm.}
    \label{fig:state_algo}
    \vspace{1ex}
\end{figure}

\vspace{0.2em}
\noindent
\textbf{Selecting eviction candidates.}
To determine the best eviction candidate, we holistically estimate the benefit and cost of each eviction candidate.
To quantify the eviction benefit, we define the \textit{GPU memory pressure} as the total size of non-evicted tensors in GPU memory at time $T$. 
If the pressure exceeds the GPU memory capacity at any time, hardware page faults will occur and harm the performance. 
Thus, the eviction candidate is beneficial if it reduces the memory pressure exceeding the capacity, as shown in the shaded area in Figure~\ref{fig:state_algo}(2).
The area of the shaded area quantifies the benefit of eviction: a larger shaded area implies a larger tensor or longer inactive period. Using this criterion, \pname{} sorts the inactive periods of all tensors to find the best eviction candidates.

The cost of an eviction candidate is quantified as the sum of eviction and prefetch latencies of this tensor. Thus, \pname{} favors migrations with low migration latencies, as other migrations may exclusively occupy the interconnect for a longer time and cause contentions, as shown in Figure~\ref{fig:state_algo}(3).

For example, in Figure~\ref{fig:state_algo}(1), we designate the best candidate as tensor $X$'s inactive period $A$. By evicting $X$ during $A$, we reduce the most pressure over the capacity limit (i.e., largest shaded area \circlepblue{3} in Figure~\ref{fig:state_algo}(2)) while causing the least I/O bandwidth overhead (shown as \circlepblue{1} + \circlepblue{2} in Figure~\ref{fig:state_algo}(3)).
Specifically, it has highest benefit-cost ratio of $\circlepblue{3} / (\circlepblue{1} + \circlepblue{2})$.


\vspace{0.2em}
\noindent
\textbf{Choosing eviction destination.}
After selecting the candidate tensors, we need to decide between two potential migration destinations, SSD and host memory, as they provide different capacities and bandwidths.  
In \pname{}, we always attempt to evict tensors to the SSD first, due to its large capacity. In contrast, host memory only offers a limited memory capacity, and thus na\"ivly evicting to host memory easily consumes up the capacity, and falls back to evicting to SSD eventually. 
However, in some cases, we still want to leverage the valuable host memory for our tensor migration. Compared to the SSD, 
the host DRAM offers much higher access bandwidth. Thus, we only evict a tensor to host memory, when the SSD traffic is under high pressure, as shown in line 7-17 in Algorithm~\ref{alg:eviction}. 
In this way, \pname{} exploits the large SSD capacity when its bandwidth is sufficient, and utilizes the high migration bandwidth 
of GPU-Host when the SSD bandwidth is saturated. 

\vspace{0.2em}
\noindent
\textbf{Smart Tensor Eviction Scheduling.}
We describe the end-to-end procedure of \pname{}'s smart tensor eviction scheduling algorithm in Algorithm~\ref{alg:eviction}. To generate an optimized migration plan, it iteratively searches for the best eviction candidate, until the GPU memory pressure is below the capacity limit or there are no more beneficial eviction candidates.

The algorithm tracks the three global states throughout the search process: (1) a set of inactive periods, (2) the estimated memory pressure versus time, and (3) the estimated bandwidth utilizations.
In each iteration of the algorithm, it selects one eviction candidate, chooses where to evict this tensor as described above, and updates the three states accordingly.

\subsection{Smart Tensor Prefetching}
\label{subsec:tensorprefetch}

\begin{figure}[!t]
    \centering
    \vspace{-1.5ex}
    \includegraphics[width=0.9\linewidth]{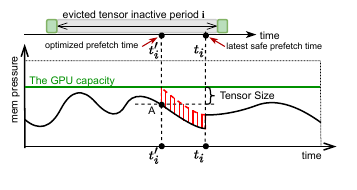}
    \vspace{-6ex}
    \caption{An example of scheduling prefetch time for one evicted inactive tensor.}
    \vspace{1ex}
    \label{fig:prefetch_opt}
\end{figure}

    

To maximize memory pressure suppression, the smart tensor eviction algorithm assumes the prefetch to be performed at the latest time that does not cause data idleness, which is defined as the \textit{latest safe prefetch time}.
However, to ensure that each prefetch completes exactly before the respective tensor turns active, the algorithm needs a perfect estimation on inactive period lengths and I/O traffic status. 
Thus, inaccurate estimation of inactive period length or I/O traffic status will incur stalls under this default prefetch policy.


Our insight is that for most DNN programs, the GPU memory pressure is under the capacity limit after scheduling the evictions. 
As shown in Figure \ref{fig:prefetch_opt}, the GPU memory pressure, presented as the black curve, is under the GPU capacity over time. The na\"ive prefetch policy does not fully utilize 
the remaining GPU memory. 

Based on our insight, \pname{} applies a smart prefetching algorithm that prefetches evicted inactive tensors eagerly to further tolerate imperfect migration decisions. \pname{} sorts all the evicted tensor inactive periods in the order of their \textit{latest safe prefetch time}. \pname{} then traverses all evicted tensor inactive periods in order and reschedules their prefetch beforehand if possible. Figure \ref{fig:prefetch_opt} shows an example. For one evicted inactive period $i$ with the latest safe prefetch time $t_i$, the algorithm searches backward from time $t_i$ until reaching the earliest time $t_i'$, when GPU can hold the entire tensor safely with the available space. In other words, the algorithm selects a time $t_i'$ at which placing this tensor on GPU will not exceed GPU memory capacity. Therefore, the algorithm schedules the prefetch for this tensor at $t_i'$, and the GPU memory pressure curve between time $t_i'$ and $t_i$ is updated. If there is not such an optimization opportunity, 
the prefetch instruction will still be scheduled at time $t_i$.

\vspace{0.2em}
\noindent
\textbf{Code Instrumentation.} To enable smart data migration, \mbox{\pname{}} utilizes deep learning comilers to automatically insert the following instructions into the 
generated GPU program: (1) \texttt{g10\_prefetch(vaddr, size)}, 
which fetches a tensor into GPU memory; (2) \texttt{g10\_pre\_evict (vaddr, size, target\_loc)}, which evicts a tensor from GPU memory to the SSD or host memory; (3) \texttt{g10\_alloc(**ptr, size)}, 
which allocates a buffer on the GPU memory asynchronously; (4) \texttt{g10\_free(*ptr)}, which frees the buffer asynchronously. We show an example of instrumented GPU program 
in Figure~\mbox{\ref{fig:code_instrumentation}}. We will further discuss these instructions in $\S$\mbox{\ref{subsec:runtime}}.

\definecolor{commentgreen}{rgb}{0,0.6,0}
\definecolor{gray}{rgb}{0.5,0.5,0.5}
\definecolor{mauve}{rgb}{0.58,0,0.82}
\definecolor{darkgreen}{rgb}{0,0.29979,0}
\definecolor{darkblue}{rgb}{0,0,0.09979}
\lstdefinestyle{CStyle} {frame=tb,
  language=C,
  aboveskip=3mm,
  belowskip=3mm,
  showstringspaces=false,
  columns=flexible,
  basicstyle={\scriptsize\ttfamily},
  numbers=left,
  numbersep=5pt,
  numberstyle=\tiny\color{gray},
  keywordstyle=\color{blue},
  commentstyle=\color{commentgreen},
  stringstyle=\color{mauve},
  breaklines=true,
  breakatwhitespace=true,
  tabsize=4,
  classoffset=0,
  morekeywords={g10_alloc, g10_free, g10_prefetch, g10_pre_evict},
  keywordstyle=\color{blue},
  classoffset=1,
  morekeywords={tensor2914}, 
  keywordstyle=\color{darkgreen},
  classoffset=2,
  morekeywords={tensor22},
  keywordstyle=\color{red},
  classoffset=3,
  morekeywords={tensor23},
  keywordstyle=\color{orange},
  classoffset=4,
  morekeywords={tensor20},
  keywordstyle=\color{cyan},
  classoffset=5,
  morekeywords={forward_conv2d_l4},
  keywordstyle=\color{violet},
}

\begin{figure}[t]
    \centering
    \vspace{-3.5ex}
\begin{lstlisting}[style=CStyle,title={\hl{}},escapechar=~~,label=code:instrumentation]
...
g10_alloc(tensor20, 40960);
g10_prefetch(tensor23, 40960);
// Kernel 2 ReLU(input, output)
forward_ReLU_l2(tensor5, tensor5);
...
g10_alloc(&tensor22, 77073360);
g10_alloc(&tensor2914, 4110417920);
// Kernel 3 MaxPool2d(input, output)
forward_MaxPool2d_l3(tensor5, tensor20);
...
// Kernel 4 Conv2d(input, output, filter, workspace)
forward_conv2d_l4(tensor20, tensor22, tensor23, tensor2914);
g10_free(tensor2914);
g10_pre_evict(tensor23, 40960, SSD);
...
// Kernel 5 BatchNorm2d(...) ~~\tikz[remember picture]{\coordinate (P1) at (-2em,-1ex)}~~
forward_BatchNorm2d_l5(tensor22, tensor28, tensor38, tensor39, tensor30, tensor31, tensor32, tensor33);
...
\end{lstlisting}
    \vspace{-2.5ex}
    \caption{An example of instrumented GPU program.}
    \vspace{1.5ex}
    \label{fig:code_instrumentation}
\end{figure}

\subsection{Unified GPU Memory and Storage}
\label{subsec:unifiedmemory}
The diversified memory and storage hierarchy (i.e., GPU memory, host memory, and SSD) inevitably increases the complexity 
of GPU memory management, and makes it challenging for \pname{} to track the memory locations for each tensor. To address this challenge, 
we develop a unified memory space. Therefore, \pname{} can plan the tensor migration 
schemes using virtual addresses, the runtime system will rely on the unified virtual memory to conduct the address translation, and 
identify the physical locations of tensors transparently. 

Prior studies~\cite{FlashMap,flatflash:asplos2019} proposed the unified address translation 
for memory-mapped SSDs, which combines the address mapping of SSDs into the page table of the virtual memory. Therefore, the page
table entries can directly point to the physical flash addresses. Although GPU provides the unified virtual memory (UVM) to manage the host memory and GPU memory in a
unified space~\cite{gpupaging, mosaic, fpgaaddr:hpca2017, batchuvm:asplos2020},
current GPU UVM does not support flash memory. 

\mbox{\pname{}} integrates the GPU memory, host memory, and flash memory into a unified memory space for enabling transparent tensor migrations. 
With unified memory, all tensors are managed at the regular 4KB page granularity. For the tensors whose size is less than 4KB, 
\mbox{\pname{}} will compact them in a page to minimize the memory fragmentation across different memory types. 
As the GPU and host interact with the SSD at the regular page granularity, the I/O amplification of the SSD will not be worse than commodity SSDs.

With the UVM extension, \mbox{\pname{}} has a unified address translation layer in the memory manager, where the flash address mappings 
in the flash translation layer have been integrated into the page table of the GPU UVM.
In this case,
the page table entry (PTE) will either point to an address in host memory or GPU memory or flash memory. 
\mbox{\pname{}} allows the SSD controller to update the page table entries (PTEs), 
when garbage collection (GC) of the SSD moves valid flash pages to a new flash block. \mbox{\pname{}} relies on the existing 
UVM supports to maintain the consistency of the host-side unified page table and GPU-side local page table, as well as the TLBs. 
As \mbox{\pname{}} migrates tensors among GPU memory, host memory, and SSD at page granularity, 
the corresponding PTEs and TLBs will also be updated with the new page address.
Since the PTE and its corresponding TLB are always updated, the unified memory system handles  
address translations and paging to load data from SSD or host memory to the GPU memory.

The UVM extension simplifies the programmability and enables transparent tensor migration. 
Its page fault handling mechanism may incur extra performance overhead. However, the smart tensor migration mechanism 
in \mbox{\pname{}} minimizes unexpected page faults and data migrations, which makes the UVM extension an appealing feature 
(see Figure~\mbox{\ref{fig:e2e_throughput}}).

\subsection{Tensor Migration with Extended UVM}
\label{subsec:runtime}

\begin{figure}[!t]
    \centering
    \includegraphics[width=1.0\columnwidth]{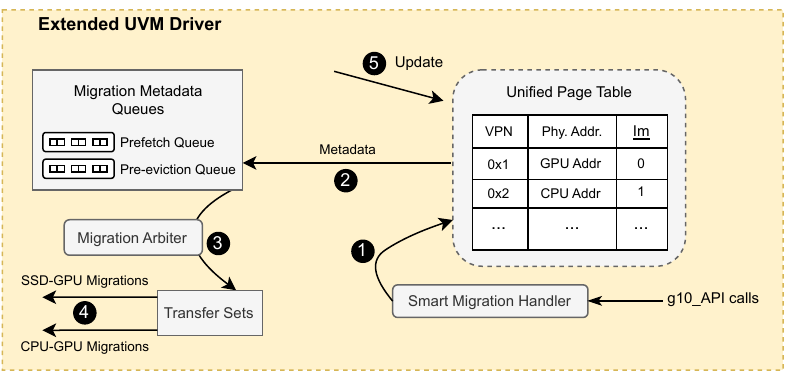}
    \caption{The workflow of runtime migrations in \pname{}.}
    \vspace{1ex}
    \label{fig:sysarch_support}
    \vspace{2ex}
\end{figure}



\pname{} supports smart tensor migration with the extended UVM ($\S$\ref{subsec:unifiedmemory}). It extends the device UVM driver to implement the 
smart migration handler on the host. We show the workflow of tensor migration in Figure~\ref{fig:sysarch_support}. 
Upon executing \texttt{g10\_pre\_evict(vaddr, size, target\_loc)}, CUDA runtime will 
send an exception to the migration handler on the host side. The migration handler will initiate the migration of the corresponding tensor, and 
migrate the tensor to the specified location \texttt{target\_loc} via the DMA engine. Note that \pname{} will rely on the unified memory system for the address translation for \texttt{vaddr}, and use the \texttt{size} to decide how many pages it will migrate. 
Upon executing \texttt{g10\_prefetch(vaddr, size)}, the tensor migration handler will access the unified memory with \texttt{vaddr}. It will initiate 
the prefetching process and request the GPU DMA engine to fetch the tensor from the host memory or SSD. 


As shown in Figure~\ref{fig:sysarch_support}, 
for tensor evictions and prefetching, \pname{} will rely on the unified page table to identify the physical locations of tensors (\circlep{1}). For pre-evictions, \pname{} will look up the GPU page to be evicted. 
After that, the migration metadata will be stored in corresponding Migration Metadata Queues (\circlep{2}). The Migration Arbiter will select several page migrations to form the next migration batch and store them in the Transfer Sets (\circlep{3}). During this procedure, The \pname{} driver will also communicate with GPU to allocate GPU memory on demand. The migrations in the Transfer Sets will be batched periodically, the corresponding SSD-GPU data transfer will be handled by the Direct Storage Access (DSA) process, and CPU-GPU data transfer will be handled by the DMA process (\circlep{4}). After the data migrations, 
the unified page table and corresponding TLB entries will be updated (\circlep{5}).

\pname{} fully utilizes the GPU-Host bandwidth and storage bandwidth with data batching. 
Migration Arbiter applies different priorities to different migration queues (e.g., page faults have the highest priority). 
\pname{} will calculate the batch number in the next round to fully saturate the bandwidth.




\section{Implementation details}
\label{sec:implementation}


\vspace{0.2em}
\noindent
\textbf{Tensor vitality analyzer.} The tensor vitality analyzer is a static analysis tool, which is compatible with the deep learning compiler PyTorch. 
The analyzer takes a DNN model and the profiled execution time of each kernel as inputs. After the static analysis ($\S$\ref{subsec:tensorAnalysis}), 
it generates instrumented CUDA programs. We take the instrumented program into the simulator framework (see below) to simulate the entire \pname{} system. 



\vspace{0.2em}
\noindent
\textbf{Simulator framework.}
To efficiently simulate the executions of diverse DNN models, we first run these real models on a real A100 GPU and trace the execution of all kernels. We build a simulation framework based on UVMSmart\mbox{\cite{InterplayUVM}} and GPGPU-Sim\mbox{\cite{gpgpusim}} to simulate the UVM, including the GPU page fault handling, data migration, and address translation. Our simulator supports taking the execution traces as input, so it can replay the kernel traces. we believe our simulation framework reasonably models the actual execution of DNN models, especially considering it replays real kernel traces collected on a real GPU.


We focus on the address translation and coherency support for the unified page tables. We modeled the latency overheads caused by the host page fault handler, 
the interaction between the GPU and CPU for the page fault handler, and page table walks, inside our timing model for accurate measurements.

When incorporating SSD into the UVM system, we follow the approach described in prior studies~\cite{flatflash:asplos2019}. 
We rely on the host page fault handing mechanism to do the address translation. Upon access to pages that do not reside in GPU memory, 
the GPU page fault handler will raise an interrupt to the host, and the host is responsible for moving data. 
To simulate the SSD internals and capture their activities, such as garbage collection (GC) and flash chip accesses, in our evaluation, 
we developed an SSD simulator based on SSD-Sim\mbox{\cite{ssdsim}} and integrated it 
into our simulator framework. Therefore, as we measure the overall system performance during the experiments, the internal SSD activities are considered.

\bgroup
\setlength\tabcolsep{3pt}
\begin{table}[t]
\centering
\caption{Evaluated DNN models and datasets.}
\vspace{-1ex}
\small
\begin{tabular}{|l|l|l|l|}
\hline
\textbf{Model}             & \textbf{\# Kernels} & \textbf{Source}           & \textbf{Dataset}  \\ \hline
BERT~\cite{devlin2018bert}  & 1368       & Hugging Face     & CoLA \\ \hline
ViT~\cite{vit}              & 1435       & Hugging Face     & ImageNet     \\ \hline
Inceptionv3~\cite{szegedy2016rethinking}       & 740       & Pytorch Examples & ImageNet \\ \hline
ResNet152~\cite{ResNet}         & 1298      & Pytorch Examples & ImageNet \\ \hline
SENet154~\cite{SENet}          & 2318      & Pytorch Examples & ImageNet \\ \hline
\end{tabular}
\label{table:benchmarks}
\end{table}
\egroup

\begin{table}[t]
\centering
\caption{System Configuration.}
 \vspace{-1ex}
\small
\begin{tabular}{|l|l|}
\hline
\textbf{CPU Main Memory}                 & 128GB DDR4                   \\ \hline
\textbf{GPU}                             & NVIDIA A100                  \\ \hline
\textbf{GPU Memory}                      & 40GB HBM2e                   \\ \hline
\textbf{Page Size}                       & 4KB                          \\ \hline
\textbf{SSD Read/Write Bandwidth}      & 3.2/3.0 GB/s                      \\ 
\hline
\textbf{SSD Read/Write Latency}                     & 20/16 $\mu$s                     \\ 
\hline
\textbf{SSD Capacity}      & 3.2 TB                      \\ 
\hline
\textbf{Interconnect}                    & PCIe Gen3 x16                \\ \hline
\textbf{GPU Page Fault Handling Latency} & 45 $\mu$s                     \\ \hline
\end{tabular}
\label{table:system_configuration}
\vspace{4ex}
\end{table}

\section{Discussion and Future Work}
\label{sec:discussion}

\noindent\textbf{Multi-GPU support}. G10 can be simply extended to effectively support multiple GPUs for three reasons. First, as multiple GPUs share SSDs, and each GPU can run independently, we can deploy the smart tensor migration mechanism of G10 on each GPU. Therefore, each GPU will make 
its own decisions on the tensor migrations. Second, current UVM has supported multiple GPUs, which has created a unified memory space across the host memory 
and all \mbox{GPUs'} memory. The UVM extension of G10 supports multiple GPUs by integrating the shared flash memory space into the existing UVM as discussed in \S\mbox{\ref{subsec:unifiedmemory}}. Third, as we increase the number of GPUs, we may want to increase the number of SSDs for increasing aggregated storage bandwidth. Since the SSD array (e.g., using RAID) is shared by multiple GPUs, G10 will treat the SSD array as a shared flash memory space and integrate it into the UVM. Our evaluation (\S\mbox{\ref{sec:eval_ssd_bw}}) will conduct the sensitivity analysis as we increase the number of SSDs. We wish to explore the multi-GPU support as 
future work.

\section{Evaluation}
\label{sec:eval}


We show that (1) \pname{} outperforms state-of-the-art designs by up to 1.75$\times$ for training large DNN models that exceed GPU on-board memory capacity ($\S$\ref{sec:eval_overall}); 
(2) \pname{} supports larger batch sizes with better performance than other designs ($\S$\ref{sec:eval_batch_size}); 
(3) \pname{} saves host memory capacity with negligible performance degradation ($\S$\ref{sec:eval_host_mem}); 
(4) \pname{} improves DNN training performance with different hardware settings ($\S$\ref{sec:eval_ssd_bw}); 
(5) \pname{}'s scheduling algorithm is resilient against profiling errors ($\S$\ref{sec:eval_profiling_error}); 
(6) \mbox{\pname{}} has negligible negative impact on the SSD lifetime (\mbox{$\S$\ref{subsec:ssd_lifetime}}).

\begin{figure}[t]
    \centering
    \includegraphics[width=\linewidth]{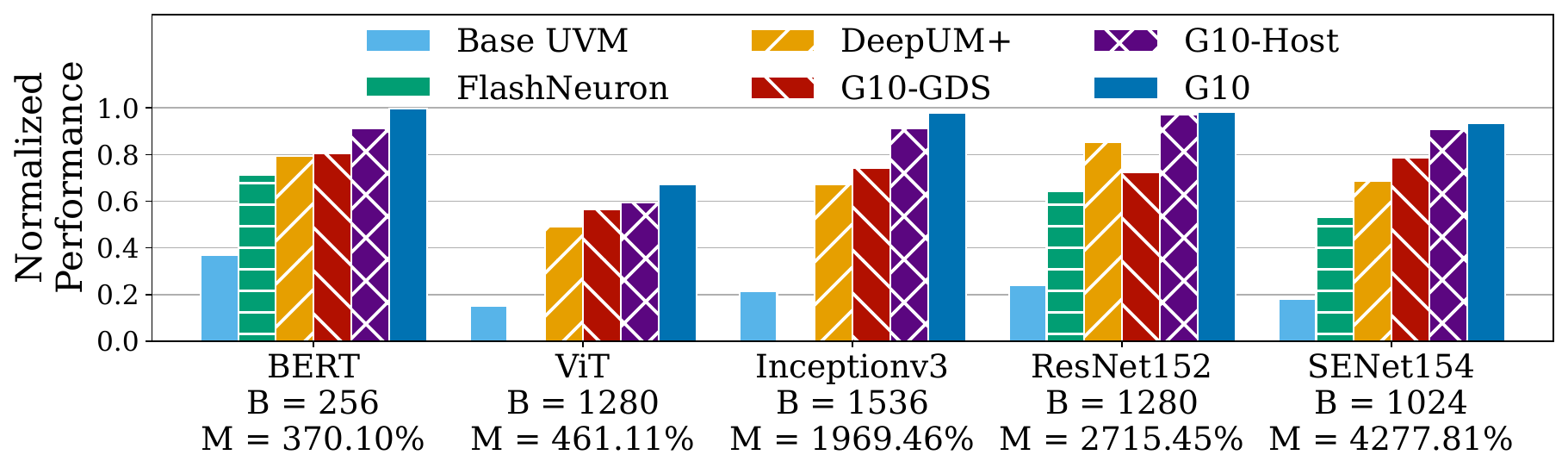}
	\caption{DNN training throughput normalized to the ideal performance. \texttt{B} is batch size. \texttt{M} is the total memory 
	consumption of the DNN w.r.t. GPU memory capacity.
	}
    \label{fig:e2e_throughput}
     \vspace{5ex}
\end{figure}

\subsection{Experimental Setup}



We evaluate \pname{} with diverse DNN models in Table~\ref{table:benchmarks}, including transformer-based models (BERT and ViT) and CNNs (ResNet, Inceptionv3, and SENet). The models are retrieved from PyTorch examples~\cite{pytorchexamples} and the Hugging Face public repositories~\cite{huggingface}, and the training datasets include CoLA~\cite{warstadt2018neural} and ImageNet~\cite{krizhevsky2012imagenet}. We use FP32 format for the tensor representation.
We vary the batch size for each model to study the impact of different memory demands.


\vspace{0.2em}
\noindent
\textbf{System configuration.}
Table~\ref{table:system_configuration} shows the hardware configuration of our experimental testbed. We set the SSD parameters based on Samsung Z-NAND SSD~\cite{sumsung-sz985}. The host memory, GPU, and SSD are connected with a PCIe interconnect that can deliver a bandwidth of 15.754 GB/s bidirectionally.
We model the UVM system following prior works~\cite{InterplayUVM, towardsUVM}.

\begin{figure}[!tph]
    \centering
    \includegraphics[width=\linewidth]{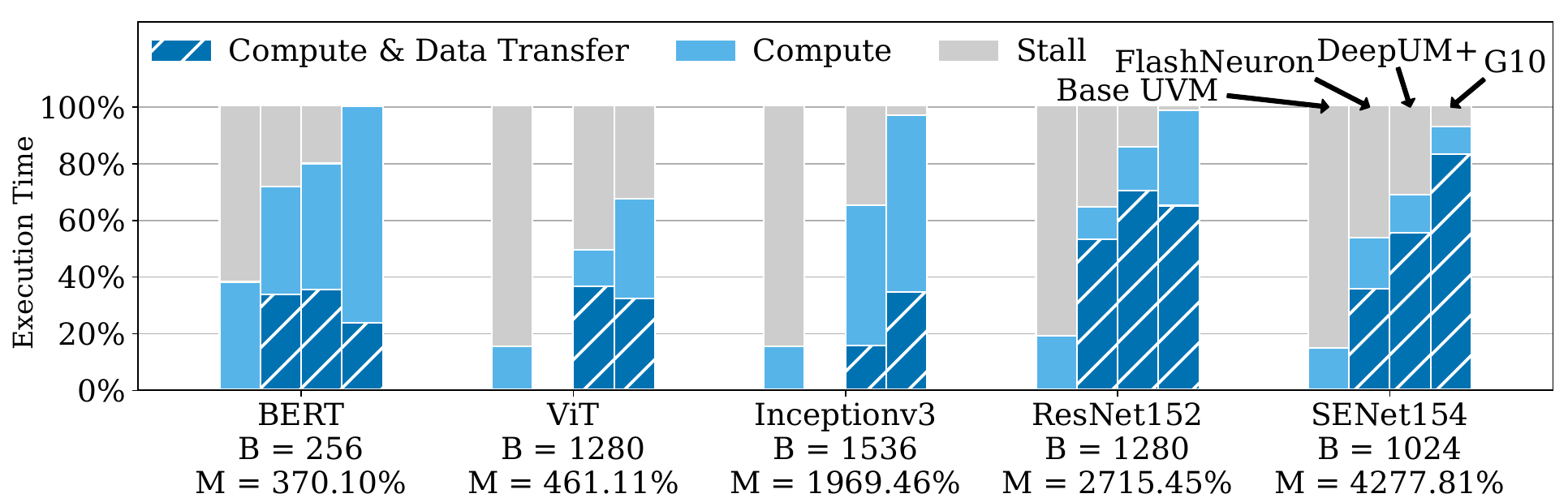}
    \caption{Execution time breakdown of training (left to right: Base UVM, FlashNeuron, DeepUM+, \pname{}).}
    \label{fig:e2e_breakdown}
    \vspace{3ex}
\end{figure}

We compare \pname{} with several state-of-the-art GPU memory-expanding solutions: DeepUM+\cite{deepum} and FlashNeuron\cite{flashneuron}. We also evaluate \pname{} with different host memory capacities. 
As hardware capabilities evolve over time, we conduct sensitivity analysis with different SSD bandwidths. 
To summarize, we compare \pname{} against the following baseline designs:
\vspace{0.2em}
\begin{itemize}[leftmargin=*,nosep]
    \vspace{1ex}
    \item \textbf{Ideal}: a GPU with infinite on-board memory, which gives the theoretically best performance.
    \vspace{1ex}
    \item \textbf{Base UVM}: the basic GPU-CPU-SSD UVM system with only on-demand page migrations via page faults. 
    \vspace{1ex}
    \item \textbf{DeepUM+}: a UVM system using a correlation-based prefetcher to prefetch data to the GPU memory. We extend the original GPU-CPU-based DeepUM design~\cite{deepum} to support SSDs. Upon a GPU page eviction, if the CPU memory is full, DeepUM+ can still evict the page to the SSD.
    \vspace{1ex}
    \item \textbf{FlashNeuron}~\cite{flashneuron}: a DNN training library using direct GPU-SSD communication to selectively swap intermediate tensors (instead of all tensors) to the SSD. Since FlashNeuron worked in a traditional non-UVM style, we used FlashNeuron's memory manager for fair comparison.
    \vspace{1ex}
\end{itemize}

\begin{figure*}[t]
    \centering
    \vspace{-0ex}
    \includegraphics[width=\linewidth]{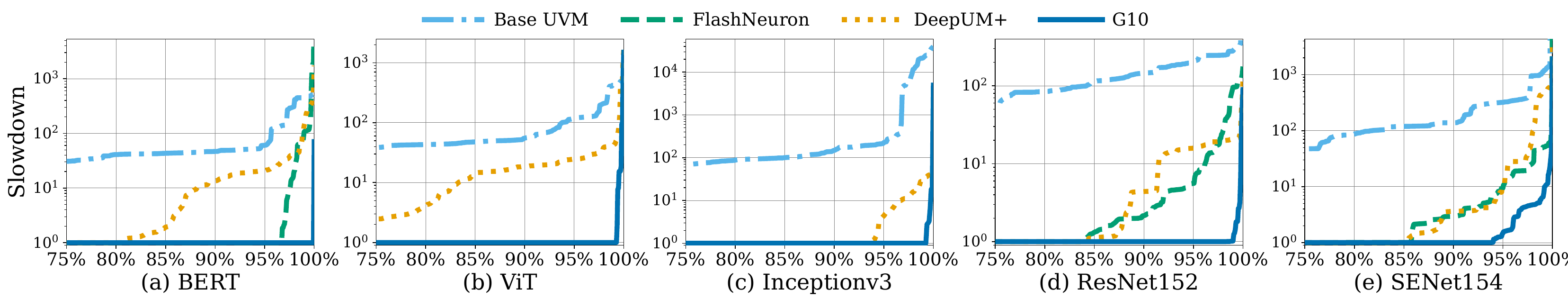}
    \caption{Distribution of kernel execution time slowdown normalized to ideal performance (lower is better).}
    \label{fig:kernel_time_cdf}
    \vspace{-1ex}
\end{figure*}

\subsection{End-to-end Performance}\label{sec:eval_overall}

We show the end-to-end DNN training throughput of different benchmarks in 
Figure~\ref{fig:e2e_throughput}\footnote{FlashNeuron fails to execute ViT and Inceptionv3 models when their batch size is large, as 
the GPU memory cannot host all the tensors required for a kernel execution, due to the limited GPU memory capacity. 
}
On average, \pname{} outperforms FlashNeuron by 1.56$\times$ and DeepUM+ by 1.31$\times$. Compared to the ideal system with infinite GPU memory, \pname{} unleashes 90.3\% of the ideal performance using limited GPU memory. 

\vspace{0.2em}
\noindent
\textbf{DNN training throughput.}
As shown in Figure~\ref{fig:e2e_throughput}, Base UVM performs 4.55$\times$ worse than the ideal, due to the significant page fault overhead.
With heuristic-based tensor eviction and prefetching, FlashNeuron and DeepVM+ improve the performance over Base UVM by 2.46$\times$ 
and 3.12$\times$, respectively. However, both of them are still much slower than the ideal performance. 
Although DeepUM+ supports DNN models with large memory demands, its correlation-based prefetching mechanism cannot capture rich DNN semantics.

\pname{} outperforms FlashNeuron and DeepVM+ by up to 1.75$\times$, which demonstrates the effectiveness of the smart tensor migration algorithm 
in capturing DNN semantics. 
For most benchmarks, \pname{} achieves nearly ideal performance by exploiting the deterministic 
dataflow of DNN workloads and best utilizing the limited I/O bandwidth. The only exception is ViT, which has high migration I/O bandwidth 
demand when the batch size is large.

To further understand the benefits of \mbox{\pname{}}, we gradually enable the features of \mbox{\pname{}}. Therefore, we have 
(1) \mbox{\textbf{\pnameGDS{}}} that only supports tensor migrations between GPU and SSD; 
(2) \mbox{\textbf{\pnameGDSFull{}}} that enables tensor migrations among GPU, host, and SSD; and (3) \mbox{\textbf{\pname{}}} that extends \mbox{\textbf{\pnameGDSFull{}}} by having 
the UVM extension which unifies the GPU memory, host memory, and SSD ($\S$\mbox{\ref{subsec:unifiedmemory}}). 
As shown in Figure~\mbox{\ref{fig:e2e_throughput}}, \mbox{\pnameGDS{}} outperforms existing solutions for most DNN workloads, because of 
its smart tensor migrations. \mbox{\pnameGDSFull{}} further improves the performance as it utilizes the host memory. For ResNet152 workload, 
\mbox{\pnameGDS{}} does not perform better than DeepUM+, because \mbox{\pnameGDS{}} can only migrate tensors between GPU and SSD. However, by enabling tensor migrations between GPU and host, 
\mbox{\pnameGDSFull{}} outperforms DeepUM+ by 1.23$\times$. With UVM extension enabled, \mbox{\pname{}} further improves the performance, due to the reduced software overhead of accessing flash pages and handling page faults.

\vspace{0.2em}
\noindent
\textbf{Execution time breakdown.}
The performance benefit of \pname{} comes from the better overlapping between computation and memory swapping.
Figure~\ref{fig:e2e_breakdown} shows the percentage of time during which tensor migrations perfectly overlap with GPU computation, 
and the percentage of time when tensor migrations stall GPU computation.
Compared to all other designs, \pname{} has the least stall time, since it generates a better swapping schedule than other designs.

Figure~\ref{fig:kernel_time_cdf} further shows how many kernels are stalled by tensor swapping. For Base UVM, more than half of the kernels (truncated in the figure) suffer page fault overhead. FlashNeuron and DeepUM+ reduce the number of affected kernels, but both designs still cause significant slowdown to many kernels (4\%--30\% of kernels). 
With \pname{}, only 1\%--6\% of kernels perform worse than the ideal case.

\begin{figure}[t]
    \centering
    \includegraphics[width=\linewidth]{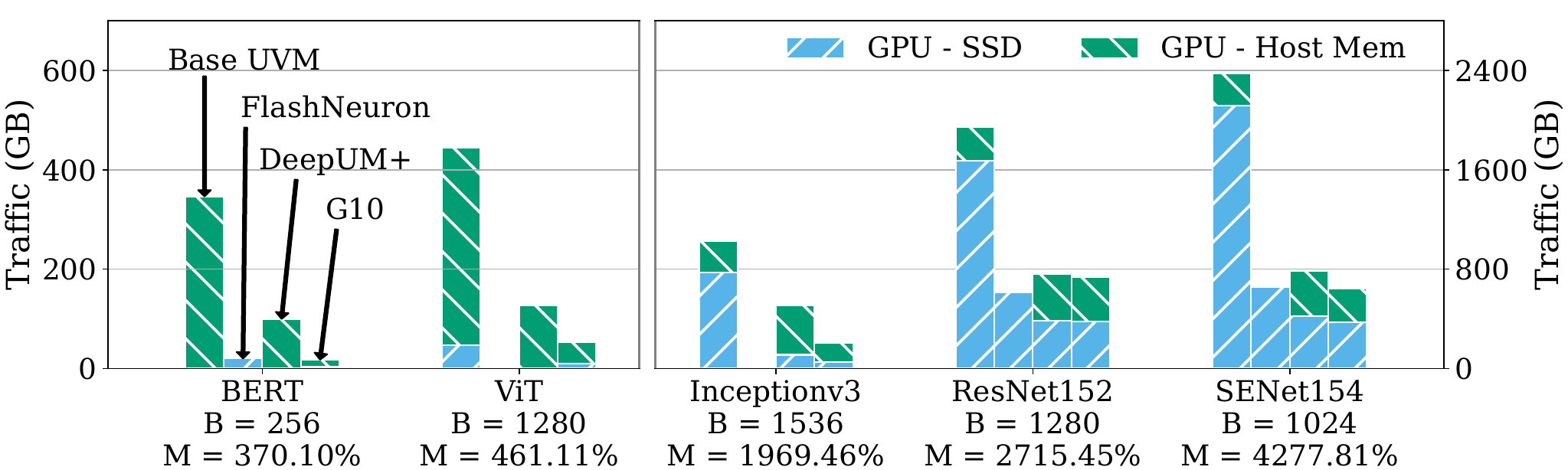}
    \caption{Tensor migration traffic breakdown.}
    \label{fig:swap_traffic}
    \vspace{1ex}
\end{figure}

\begin{figure*}[t]
    \centering
    \vspace{-1ex}
    \includegraphics[width=\linewidth]{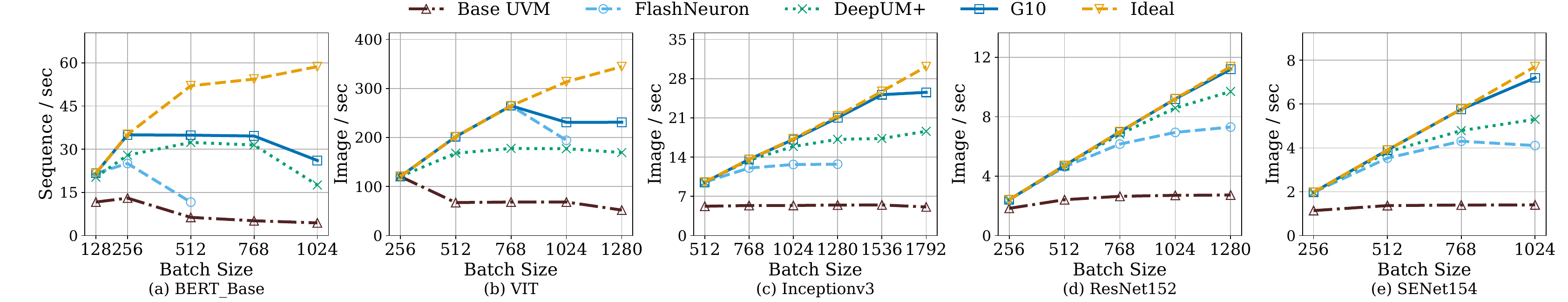}
    \vspace{-4.5ex}
    \caption{Training throughput with varying batch sizes.}
    \vspace{-2ex}
    \label{fig:eval_batch_size}
\end{figure*}

\begin{figure*}[t]
    \centering
\includegraphics[width=\linewidth]{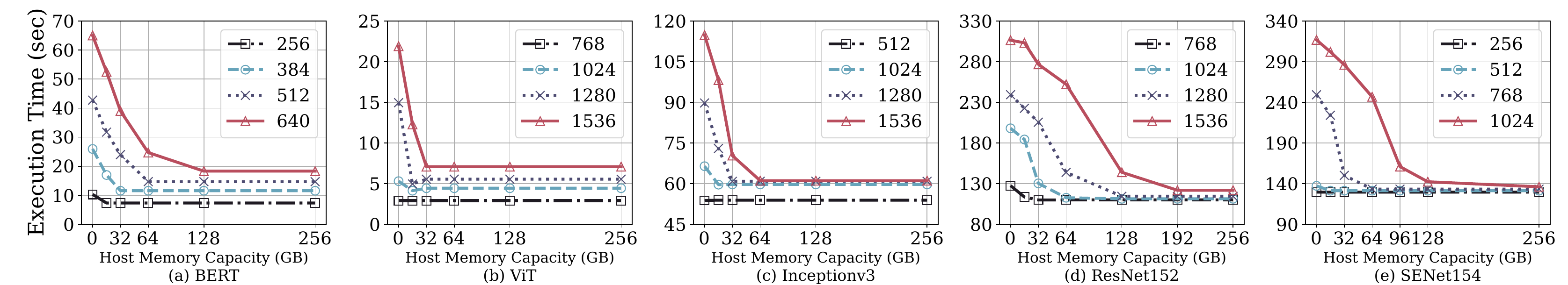}
    \vspace{-4.5ex}
    \caption{Execution time as we vary the host memory capacity.}
    \label{fig:eval_host_mem_size}
    \vspace{-2ex}
\end{figure*}

\vspace{0.2em}
\noindent
\textbf{Tensor migration traffic.}
To understand how \pname{} utilizes the available I/O bandwidth, we show the total migration traffic of GPU-SSD and GPU-Host in Figure~\ref{fig:swap_traffic}.
Due to the inefficiency of heuristic-based migration policies (e.g., LRU policy and linear selection\mbox{\cite{flashneuron}}), \BU and \DU schedule more tensor evictions than necessary. 
On the contrary, \FN does not schedule a sufficient number of evictions as it does not swap weight tensors, so it cannot reserve enough space for future tensors in a timely manner.

We also observe that a small amount of host memory plays a critical role for \pname{} to tolerate tensors that have high migration bandwidth demands. 
Particularly, transformer models (BERT and ViT) are more bandwidth-intensive, so \pname{} directs most of their migration traffic to the host memory. 
CNN models are more compute-intensive, thus, the SSD bandwidth can sustain more than half of the migration traffic. By fully utilizing the available bandwidth, 
\pname{} unleashes the potential of the GPU-CPU-SSD unified memory.




\subsection{Performance with Varying Batch Size}\label{sec:eval_batch_size}

As batch size varies, the performance of \pname{} is always the closest to ideal among all the designs. 
In Figure~\ref{fig:eval_batch_size}, while most designs achieve the ideal performance when the batch sizes are small and the memory demand is low, 
\pname{} can tolerate larger batch sizes and higher memory demands.
With larger batch sizes, more tensors must be swapped with the limited I/O bandwidth.
Thus, it is more crucial to make smart migration decisions to hide the migration latency and avoid stalling future kernels.
Despite significantly outperforming \mbox{\BU}, \mbox{\DU}, and \mbox{\FN} quickly fall behind the ideal performance as batch size increases, 
due to the sub-optimal swapping policies.
\pname{} still timely delivers required data to the active kernels under strict capacity and bandwidth limitations in most cases, thanks to its intelligent tensor migrations. In general, \mbox{\pname{}} outperforms \mbox{\FN} and \mbox{\DU} by up to 2.67$\times$ and 1.45$\times$, respectively.

As batch size continues to increase, the performance of all designs eventually degrades, but \pname{} still outperforms all other designs. 
If the total memory consumption of the current and the next kernel exceeds GPU memory capacity, data required by the next kernel cannot be ready in GPU before the kernel starts.
Thus, the next kernel inevitably stalls due to poor overlapping between computation and data transfer.

\subsection{Impact of Varying Host Memory Capacity}\label{sec:eval_host_mem}

\begin{figure}[t]
    \centering
    \vspace{1.5ex}
    \includegraphics[width=\linewidth]{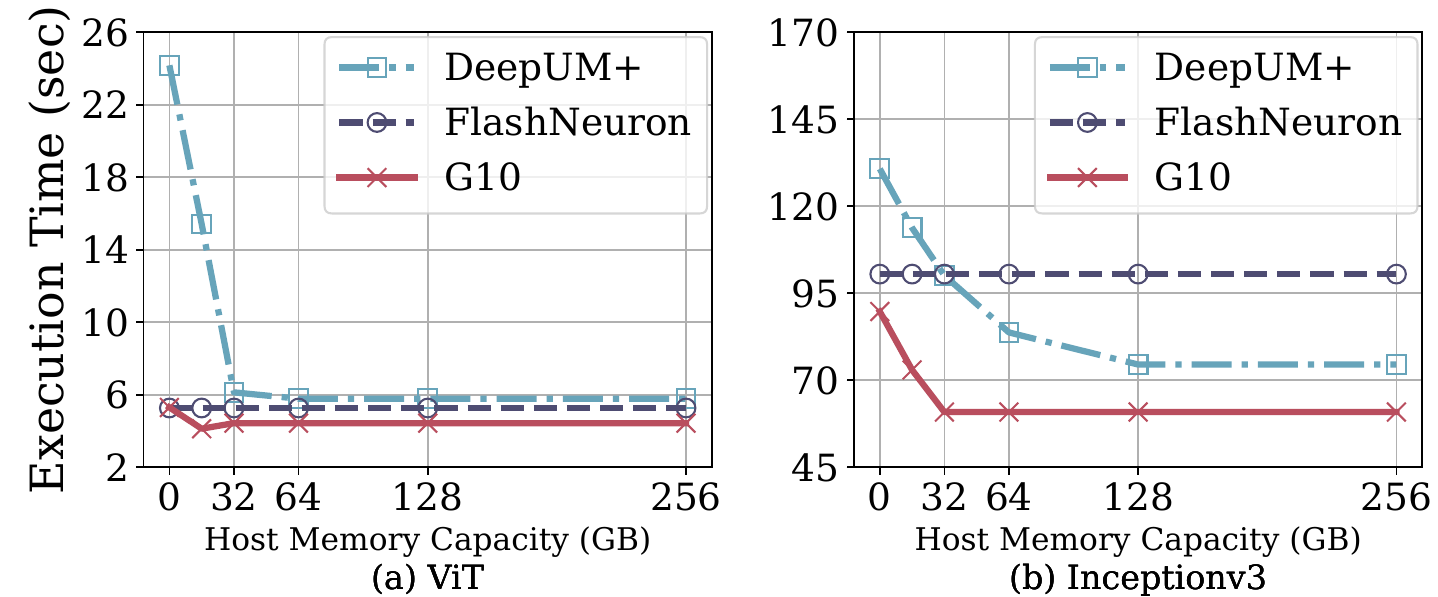}
    \vspace{-3.5ex}
    \caption{Performance comparison of \pname{}, DeepUM+, and FlashNeuron with different host memory capacity.}
    \vspace{1.5ex}
    \label{fig:appendix_host_mem_size}
\end{figure}

While using the cost-efficient SSD to expand GPU memory capacity, \pname{} also leverages the host memory bandwidth to compensate for tensors that cannot be swapped into and back from the SSD within their inactive periods. Since most tensors do not require high migration bandwidth (Figure~\ref{fig:inactive_distriubtion}), \pname{} only needs a small amount of host memory to tolerate them.
Figure~\ref{fig:eval_host_mem_size} shows \pname{}'s performance with different host memory capacities.
For most DNN models with small batch sizes, 32GB of host memory is sufficient for \pname{} to fully utilize the migration bandwidth between the host and GPU. The host memory capacity demand grows linearly with the batch size, as the sizes of the migrated tensors grow linearly.


As we vary the host memory capacity, we also compare \mbox{\pname{}} with \mbox{\DU} and \mbox{\FN}. We use two representative models: ViT (transformer) with the batch size of 1024 and Inceptionv3 (CNN) with the batch size of 1280. We show the results in Figure~\mbox{\ref{fig:appendix_host_mem_size}}. When there is no host memory, \mbox{\pname{}} outperforms DeepUM+ and FlashNeuron by 2.58$\times$ 
and 1.04$\times$ on average, respectively. This is because DeepUM+ relies on conventional GPU UVM and incurs a significant number of page faults. 
As we increase the host memory capacity, the performance of DeepUM+ is improved, however, 
it still performs 1.26$\times$ worse than \mbox{\pname{}}. As FlashNeuron fully relies on GPUDirect Storage and does not use host memory, its performance is barely affected as we vary the host memory capacity.  
Because of smart data migrations, \mbox{\pname{}} always performs better than FlashNeuron (1.33$\times$ on average).


\subsection{Impact of Varying SSD Bandwidth}\label{sec:eval_ssd_bw}

\begin{figure*}[!t]
    \centering
    \vspace{-2ex}
    \includegraphics[width=\linewidth]{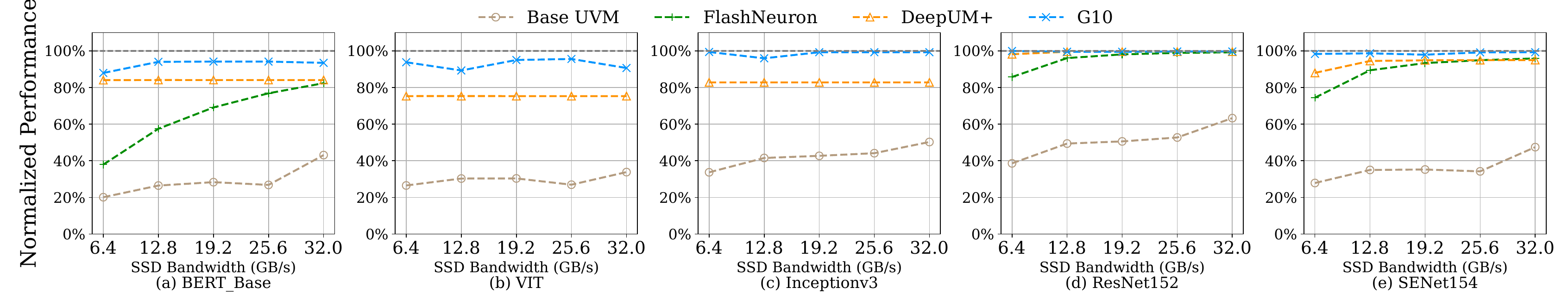}
    \vspace{-4.5ex}
    \caption{Performance with varying SSD bandwidth (normalized to ideal).}
    \label{fig:eval_vary_ssd_bw}
    \vspace{-2ex}
\end{figure*}

We now examine \pname{} with different SSD bandwidths (e.g., stacking multiple SSDs or using a higher-end SSD). 
For increased bandwidths, we assume the interconnect is PCIe 4.0 x16 (32 GB/s).
In Figure~\ref{fig:eval_vary_ssd_bw}, \pname{} outperforms all other designs regardless of the SSD bandwidth.
For all benchmarks, 1 to 4 SSDs (up to 12.8 GB/s) are sufficient for \pname{} to achieve 90\% to 100\% of the ideal performance.
BERT and ViT fail to attain the ideal performance because they are bottlenecked by the interconnect bandwidth (i.e., always swapping to host still cannot satisfy the bandwidth requirement). \pname{} exploits the high migration bandwidth of the host memory while best utilizing the SSD to reduce host memory pressure ($\S$\ref{sec:eval_host_mem}). In contrast, even with enough SSDs to saturate the interconnect bandwidth, \FN and \DU still only achieve 70\%-80\% of ideal performance for BERT and ViT.


\subsection{Impact of Profiling Errors}\label{sec:eval_profiling_error}

\begin{figure}[t]
    \centering
    \vspace{0ex}
    \includegraphics[width=\linewidth]{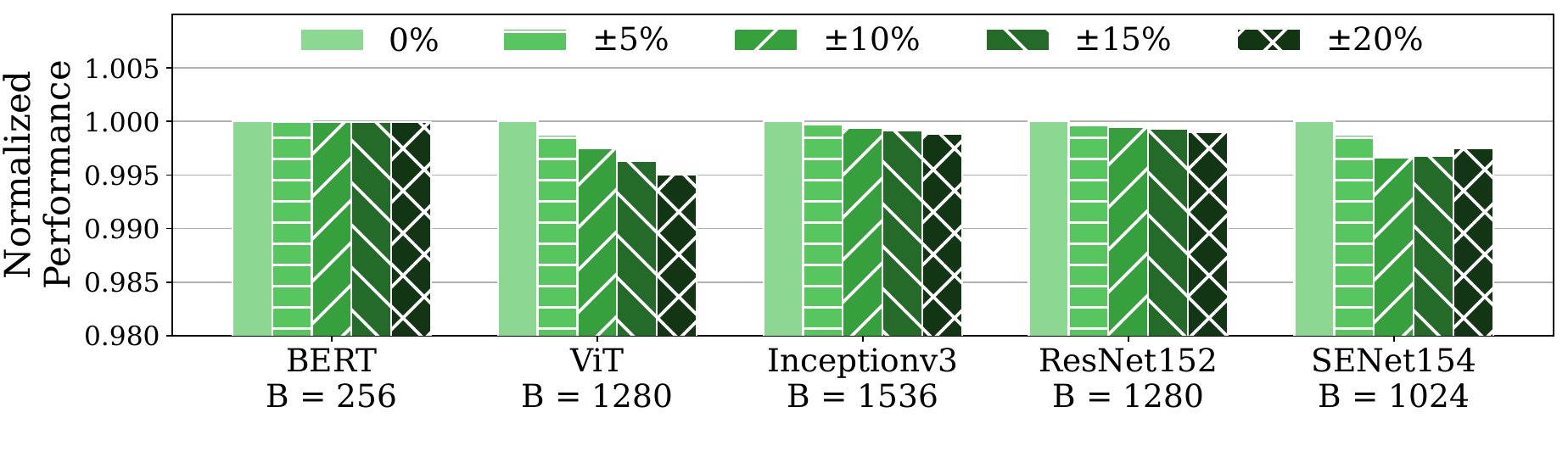}
    \vspace{-6.5ex}
    \caption{Performance of \pname{} under various degrees of kernel timing prediction errors (normalized to no error).}
    \vspace{2ex}
    \label{fig:eval_profile_noise}
\end{figure}

To understand the robustness of \pname{}'s scheduling algorithm against profiling errors, we add random noises to the execution time of each kernel in our simulator.
Figure~\ref{fig:eval_profile_noise} shows the performance of \pname{} with various degrees of profiling errors.
For all benchmarks, the performance degradation is under 0.5\% even when the profiling error is $\pm$20\%.
The profiling errors only affect the estimation of tensor inactive period lengths. 
\pname{} tolerates such errors by eagerly prefetching a tensor before it is used ($\S$\ref{subsec:tensorprefetch}). In most cases, 
the early prefetch can tolerate the profiling inaccuracy.


\subsection{Impact on SSD Lifetime}\label{subsec:ssd_lifetime}

As reported in the released datasheet of Samsung Z-SSD SZ985\mbox{\cite{sumsung-sz985}}, its device endurance is 30 Drive Writes Per Day (DWPD) for five years. 
According to our study, DNN workloads 
incur almost 50\% writes and 50\% reads. 
In this case, the SSD lifetime of \mbox{\pname{}} would be 30 DWPD * 1825 days (5 years) 
* 3.2TB / 3 GB/s * 2 = 3.7 years, when it is used continuously. Considering DNN workloads are data intensive and a commodity SSD usually lasts 3-5 years, the impact on SSD lifetime is not much of a concern. Based on Figure~\mbox{\ref{fig:swap_traffic}}, we further break down the traffic into reads and writes. \mbox{\pname{}} incurs 1.37$\times$ and 
2.20$\times$ less writes than DeepUM+ and FlashNeuron, respectively. As SSD lifetime is affected by the write traffic, \mbox{\pname{}} can achieve improved lifetime than state-of-the-art solutions.

%
%

\section{Related Work}

\noindent
\textbf{GPU memory wall}.
DNN workloads are heavily using GPUs. 
They rely on GPU memory and host memory to host their working sets. 
However, due to the limited capacity, they cannot host large 
models~\cite{assaf:sysml2019, TensorFlow2015,memorywall:micro2018,diannao,peng2018optimus}. 
An alternative approach is to bring Flash closer to GPUs, such as GPUDirect Storage~\cite{gpudirectstorage}, 
ZnG~\cite{zng:isca2020, flashgpu}, and AMD's SSG~\cite{ssg, ssg-detail}.
ZnG replaced GPU memory with flash chips and hard coded the flash addresses in the GPU MMU~\cite{zng:isca2020}.
SSG and GPUDirect Storage enable GPU to directly 
communicate with SSDs via the PCIe interface~\cite{ssg-detail}.
Unfortunately, their performance is bottlenecked by the PCIe bandwidth. In this paper, we develop 
a unified GPU memory system, 
and best utilize tensor behaviors to overcome the bottlenecks of slow memories. 


\vspace{0.2em}
\noindent
\textbf{New memory technologies.}
To overcome the memory scaling wall,
researchers have been mostly focused on developing
scalable memory technologies~\cite{mem_pcm_1, 3dstack:isca2008, jishen:cgo}. 
For instance, 
HBM~\cite{hbm, amd-hbm} was produced to meet the bandwidth requirement
of accelerators, but their capacity is still limited. 
Intel released
its Optane persistent memory~\cite{intel:xpoint}, and
Samsung released its ultra-low latency SSDs~\cite{znand}. \pname{} is compatible with 
the new and emerging memory and storage devices, it leverages low-cost memories to scale the GPU memory while 
reaching near-to-ideal performance. 


\vspace{0.2em}
\noindent
\textbf{Unified memory and storage.}
Prior studies showed that SSDs can be used as memory via memory-mapped 
interface~\cite{FlashMap, flatflash:asplos2019, FlashVM, namelesswrite:fast2012, 
shapeshift, DFS, gordon}.
However, they were designed for CPU-centric computing and cannot be directly applied to GPUs.
NVIDIA and AMD have been supporting UVM in their GPU products by enabling unified memory between the host and GPU~\cite{uvm, amduvm}. \pname{} advances the architecture and integrates flash 
memories into the unified memory space. To optimize data movements between the host and GPU memory, prior 
studies~\cite{SwapAdvisor,InterplayUVM,sentinel,autotm,deepum,zero_infinity} explored data localities of DNN workloads. 
\pname{} shares the same purpose with them. 
However, different from the studies like ZeRO series\mbox{\cite{zero_infinity,ZERO,zero-offload}} that offload tensors at a coarse (DNN layer) granularity, 
\mbox{\pname{}} enables tensor migrations at page granularity, and develops an active-time-aware tensor migration scheme. 
\section{Conclusion}
\label{sec:conclusion}

We present \pname{}, a unified GPU memory and storage architecture 
for scaling deep learning workloads. \pname{} is driven by our observation that 
the predictable tensor behaviors offer sufficient opportunities for \pname{} to make smart data migrations. 
Thus, we can overlap 
the GPU computation and flash accesses. With diverse DNN training 
workloads, we show that \pname{} can achieve near-ideal performance.

\begin{acks}
We thank the anonymous reviewers for their insightful comments and feedback. This work was partially supported by NSF grant CCF-2107470, NSF CAREER Award CNS-2144796, and a grant 
from the Defense Advanced Research Projects Agency (DARPA) under the award number HR00112390029. The views, opinions and/or findings expressed are those of the author and should not be interpreted 
as representing the official views or policies of the Department of Defense or the U.S. Government.
\end{acks}
\flushcolsend

\bibliographystyle{ACM-Reference-Format}
\bibliography{refs}


\begin{thebibliography}{71}


\ifx \showCODEN    \undefined \def \showCODEN     #1{\unskip}     \fi
\ifx \showDOI      \undefined \def \showDOI       #1{#1}\fi
\ifx \showISBNx    \undefined \def \showISBNx     #1{\unskip}     \fi
\ifx \showISBNxiii \undefined \def \showISBNxiii  #1{\unskip}     \fi
\ifx \showISSN     \undefined \def \showISSN      #1{\unskip}     \fi
\ifx \showLCCN     \undefined \def \showLCCN      #1{\unskip}     \fi
\ifx \shownote     \undefined \def \shownote      #1{#1}          \fi
\ifx \showarticletitle \undefined \def \showarticletitle #1{#1}   \fi
\ifx \showURL      \undefined \def \showURL       {\relax}        \fi
\providecommand\bibfield[2]{#2}
\providecommand\bibinfo[2]{#2}
\providecommand\natexlab[1]{#1}
\providecommand\showeprint[2][]{arXiv:#2}

\bibitem[pci({[n.\,d.]})]%
        {pcisig}
 \bibinfo{year}{[n.\,d.]}\natexlab{}.
\newblock \bibinfo{title}{{PCIe 3.0 Specification}}.
\newblock \bibinfo{howpublished}{\url{https://pcisig.com/specifications}}.
\newblock


\bibitem[Abadi et~al\mbox{.}(2016)]%
        {TensorFlow2015}
\bibfield{author}{\bibinfo{person}{Martin Abadi}, \bibinfo{person}{Paul Barham}, \bibinfo{person}{Jianmin Chen}, \bibinfo{person}{Zhifeng Chen}, \bibinfo{person}{Andy Davis}, \bibinfo{person}{Jeffrey Dean}, \bibinfo{person}{Matthieu Devin}, \bibinfo{person}{Sanjay Ghemawat}, \bibinfo{person}{Geoffrey Irving}, \bibinfo{person}{Michael Isard}, \bibinfo{person}{Manjunath Kudlur}, \bibinfo{person}{Josh Levenberg}, \bibinfo{person}{Rajat Monga}, \bibinfo{person}{Sherry Moore}, \bibinfo{person}{Derek~Gordon Murray}, \bibinfo{person}{Benoit Steiner}, \bibinfo{person}{Paul~A. Tucker}, \bibinfo{person}{Vijay Vasudevan}, \bibinfo{person}{Pete Warden}, \bibinfo{person}{Martin Wicke}, \bibinfo{person}{Yuan Yu}, {and} \bibinfo{person}{Xiaoqiang Zhang}.} \bibinfo{year}{2016}\natexlab{}.
\newblock \showarticletitle{{TensorFlow: A System for Large-Scale Machine Learning}}. In \bibinfo{booktitle}{\emph{Proceedings of the 12th USENIX Symposium on Operating Systems Design and Implementation (OSDI'16)}}. \bibinfo{address}{Savannah, GA}.
\newblock


\bibitem[Abulila et~al\mbox{.}(2019)]%
        {flatflash:asplos2019}
\bibfield{author}{\bibinfo{person}{Ahmed Abulila}, \bibinfo{person}{Vikram~Sharma Mailthoday}, \bibinfo{person}{Zaid Qureshi}, \bibinfo{person}{Jian Huang}, \bibinfo{person}{Nam~Sung Kim}, \bibinfo{person}{Jin jun Xiong}, {and} \bibinfo{person}{Wen mei Hwu}.} \bibinfo{year}{2019}\natexlab{}.
\newblock \showarticletitle{{FlatFlash: Exploiting the Byte-Accessibility of SSDs within A Unified Memory-Storage Hierarchy}}. In \bibinfo{booktitle}{\emph{Proceedings of the 24th ACM International Conference on Architectural Support for Programming Languages and Operating Systems (ASPLOS'19)}}. \bibinfo{address}{Providence, RI}.
\newblock


\bibitem[Agarwal et~al\mbox{.}(2015)]%
        {gpupaging}
\bibfield{author}{\bibinfo{person}{Neha Agarwal}, \bibinfo{person}{David Nellans}, \bibinfo{person}{Mark Stephenson}, \bibinfo{person}{Mike O’Connor}, {and} \bibinfo{person}{Stephen~W. Keckler}.} \bibinfo{year}{2015}\natexlab{}.
\newblock \showarticletitle{Page Placement Strategies for GPUs within Heterogeneous Memory Systems}. In \bibinfo{booktitle}{\emph{Proceedings of the Twentieth International Conference on Architectural Support for Programming Languages and Operating Systems (ASPLOS'15)}} (Istanbul, Turkey). \bibinfo{pages}{607–618}.
\newblock


\bibitem[Agrawal et~al\mbox{.}(2008)]%
        {ssdsim}
\bibfield{author}{\bibinfo{person}{Nitin Agrawal}, \bibinfo{person}{Vijayan Prabhakaran}, \bibinfo{person}{Ted Wobber}, \bibinfo{person}{John~D. Davis}, \bibinfo{person}{Mark Manasse}, {and} \bibinfo{person}{Rina Panigrahy}.} \bibinfo{year}{2008}\natexlab{}.
\newblock \showarticletitle{{Design Tradeoffs for SSD Performance}}. In \bibinfo{booktitle}{\emph{Proceeding of the USENIX 2008 Annual Technical Conference (USENIX ATC'08)}}. \bibinfo{address}{Boston, MA}.
\newblock


\bibitem[Allen and Ge(2021)]%
        {analysis-UVM}
\bibfield{author}{\bibinfo{person}{Tyler Allen} {and} \bibinfo{person}{Rong Ge}.} \bibinfo{year}{2021}\natexlab{}.
\newblock \showarticletitle{In-Depth Analyses of Unified Virtual Memory System for GPU Accelerated Computing}. In \bibinfo{booktitle}{\emph{Proceedings of the International Conference for High Performance Computing, Networking, Storage and Analysis}} \emph{(\bibinfo{series}{SC '21})}. \bibinfo{address}{St. Louis, Missouri}.
\newblock


\bibitem[{AMD DirectGMA}({[n.\,d.]})]%
        {directgma}
\bibfield{author}{\bibinfo{person}{{AMD DirectGMA}}.} \bibinfo{year}{[n.\,d.]}\natexlab{}.
\newblock \bibinfo{howpublished}{\url{https://www.bitflow.com/technology/directgma/}}.
\newblock


\bibitem[{AMD High Bandwidth Memory}({[n.\,d.]})]%
        {amd-hbm}
\bibfield{author}{\bibinfo{person}{{AMD High Bandwidth Memory}}.} \bibinfo{year}{[n.\,d.]}\natexlab{}.
\newblock \bibinfo{howpublished}{\url{https://www.amd.com/en/technologies/hbm}}.
\newblock


\bibitem[Ausavarungnirun et~al\mbox{.}(2017)]%
        {mosaic}
\bibfield{author}{\bibinfo{person}{Rachata Ausavarungnirun}, \bibinfo{person}{Joshua Landgraf}, \bibinfo{person}{Vance Miller}, \bibinfo{person}{Saugata Ghose}, \bibinfo{person}{Jayneel Gandhi}, \bibinfo{person}{Christopher~J. Rossbach}, {and} \bibinfo{person}{Onur Mutlu}.} \bibinfo{year}{2017}\natexlab{}.
\newblock \showarticletitle{Mosaic: A GPU Memory Manager with Application-Transparent Support for Multiple Page Sizes}. In \bibinfo{booktitle}{\emph{Proceedings of the 50th Annual IEEE/ACM International Symposium on Microarchitecture (MICRO'17)}} (Cambridge, Massachusetts).
\newblock


\bibitem[Awan et~al\mbox{.}(2018)]%
        {oc-dnn}
\bibfield{author}{\bibinfo{person}{Ammar~Ahmad Awan}, \bibinfo{person}{Ching-Hsiang Chu}, \bibinfo{person}{Hari Subramoni}, \bibinfo{person}{Xiaoyi Lu}, {and} \bibinfo{person}{Dhabaleswar~K. Panda}.} \bibinfo{year}{2018}\natexlab{}.
\newblock \showarticletitle{OC-DNN: Exploiting Advanced Unified Memory Capabilities in CUDA 9 and Volta GPUs for Out-of-Core DNN Training}. In \bibinfo{booktitle}{\emph{2018 IEEE 25th International Conference on High Performance Computing (HiPC)}}. \bibinfo{pages}{143--152}.
\newblock
\urldef\tempurl%
\url{https://doi.org/10.1109/HiPC.2018.00024}
\showDOI{\tempurl}


\bibitem[Badam et~al\mbox{.}(2013)]%
        {shapeshift}
\bibfield{author}{\bibinfo{person}{Anirudh Badam}, \bibinfo{person}{Vivek~S. Pai}, {and} \bibinfo{person}{David~W. Nellans}.} \bibinfo{year}{2013}\natexlab{}.
\newblock \showarticletitle{{Better Flash Access via Shape-shifting Virtual Memory Pages}}. In \bibinfo{booktitle}{\emph{Proceedings of the First ACM SIGOPS Conference on Timely Results in Operating Systems}} \emph{(\bibinfo{series}{TRIOS '13})}. \bibinfo{address}{Farmington, PA}, Article \bibinfo{articleno}{3}, \bibinfo{numpages}{14}~pages.
\newblock
\urldef\tempurl%
\url{https://doi.org/10.1145/2524211.2524221}
\showDOI{\tempurl}


\bibitem[Bae et~al\mbox{.}(2021)]%
        {flashneuron}
\bibfield{author}{\bibinfo{person}{Jonghyun Bae}, \bibinfo{person}{Jongsung Lee}, \bibinfo{person}{Yunho Jin}, \bibinfo{person}{Sam Son}, \bibinfo{person}{Shine Kim}, \bibinfo{person}{Hakbeom Jang}, \bibinfo{person}{Tae~Jun Ham}, {and} \bibinfo{person}{Jae~W. Lee}.} \bibinfo{year}{2021}\natexlab{}.
\newblock \showarticletitle{{FlashNeuron}: {SSD-Enabled} {Large-Batch} Training of Very Deep Neural Networks}. In \bibinfo{booktitle}{\emph{19th USENIX Conference on File and Storage Technologies (FAST 21)}}. \bibinfo{publisher}{USENIX Association}, \bibinfo{pages}{387--401}.
\newblock
\showISBNx{978-1-939133-20-5}
\urldef\tempurl%
\url{https://www.usenix.org/conference/fast21/presentation/bae}
\showURL{%
\tempurl}


\bibitem[Brown et~al\mbox{.}(2020)]%
        {NEURIPS2020_1457c0d6}
\bibfield{author}{\bibinfo{person}{Tom Brown}, \bibinfo{person}{Benjamin Mann}, \bibinfo{person}{Nick Ryder}, \bibinfo{person}{Melanie Subbiah}, \bibinfo{person}{Jared~D Kaplan}, \bibinfo{person}{Prafulla Dhariwal}, \bibinfo{person}{Arvind Neelakantan}, \bibinfo{person}{Pranav Shyam}, \bibinfo{person}{Girish Sastry}, \bibinfo{person}{Amanda Askell}, \bibinfo{person}{Sandhini Agarwal}, \bibinfo{person}{Ariel Herbert-Voss}, \bibinfo{person}{Gretchen Krueger}, \bibinfo{person}{Tom Henighan}, \bibinfo{person}{Rewon Child}, \bibinfo{person}{Aditya Ramesh}, \bibinfo{person}{Daniel Ziegler}, \bibinfo{person}{Jeffrey Wu}, \bibinfo{person}{Clemens Winter}, \bibinfo{person}{Chris Hesse}, \bibinfo{person}{Mark Chen}, \bibinfo{person}{Eric Sigler}, \bibinfo{person}{Mateusz Litwin}, \bibinfo{person}{Scott Gray}, \bibinfo{person}{Benjamin Chess}, \bibinfo{person}{Jack Clark}, \bibinfo{person}{Christopher Berner}, \bibinfo{person}{Sam McCandlish}, \bibinfo{person}{Alec Radford}, \bibinfo{person}{Ilya Sutskever}, {and} \bibinfo{person}{Dario Amodei}.} \bibinfo{year}{2020}\natexlab{}.
\newblock \showarticletitle{Language Models are Few-Shot Learners}. In \bibinfo{booktitle}{\emph{Advances in Neural Information Processing Systems}}, \bibfield{editor}{\bibinfo{person}{H.~Larochelle}, \bibinfo{person}{M.~Ranzato}, \bibinfo{person}{R.~Hadsell}, \bibinfo{person}{M.F. Balcan}, {and} \bibinfo{person}{H.~Lin}} (Eds.), Vol.~\bibinfo{volume}{33}. \bibinfo{publisher}{Curran Associates, Inc.}, \bibinfo{pages}{1877--1901}.
\newblock
\urldef\tempurl%
\url{https://proceedings.neurips.cc/paper/2020/file/1457c0d6bfcb4967418bfb8ac142f64a-Paper.pdf}
\showURL{%
\tempurl}


\bibitem[Caulfield et~al\mbox{.}(2009)]%
        {gordon}
\bibfield{author}{\bibinfo{person}{Adrian~M. Caulfield}, \bibinfo{person}{Laura~M. Grupp}, {and} \bibinfo{person}{Steven Swanson}.} \bibinfo{year}{2009}\natexlab{}.
\newblock \showarticletitle{{Gordon: Using Flash Memory to Build Fast, Power-efficient Clusters for Data-intensive Applications}}. In \bibinfo{booktitle}{\emph{Proceedings of the 14th International Conference on Architectural Support for Programming Languages and Operating Systems}} \emph{(\bibinfo{series}{ASPLOS XIV})}. \bibinfo{address}{Washington, DC}, \bibinfo{pages}{217--228}.
\newblock
\urldef\tempurl%
\url{https://doi.org/10.1145/1508244.1508270}
\showDOI{\tempurl}


\bibitem[Chen et~al\mbox{.}(2014)]%
        {diannao}
\bibfield{author}{\bibinfo{person}{Tianshi Chen}, \bibinfo{person}{Zidong Du}, \bibinfo{person}{Ninghui Sun}, \bibinfo{person}{Jia Wang}, \bibinfo{person}{Chengyong Wu}, \bibinfo{person}{Yunji Chen}, {and} \bibinfo{person}{Olivier Temam}.} \bibinfo{year}{2014}\natexlab{}.
\newblock \showarticletitle{{DianNao: A Small-Footprint High-Throughput Accelerator for Ubiquitous Machine-Learning}}. In \bibinfo{booktitle}{\emph{Proceedings of the 20th International Conference on Architectural Support for Programming Languages and Operating Systems (ASPLOS'14)}}. \bibinfo{address}{Salt Lake City, UT}.
\newblock


\bibitem[Devlin et~al\mbox{.}(2018)]%
        {devlin2018bert}
\bibfield{author}{\bibinfo{person}{Jacob Devlin}, \bibinfo{person}{Ming-Wei Chang}, \bibinfo{person}{Kenton Lee}, {and} \bibinfo{person}{Kristina Toutanova}.} \bibinfo{year}{2018}\natexlab{}.
\newblock \showarticletitle{Bert: Pre-training of deep bidirectional transformers for language understanding}.
\newblock \bibinfo{journal}{\emph{arXiv preprint arXiv:1810.04805}} (\bibinfo{year}{2018}).
\newblock


\bibitem[Dosovitskiy et~al\mbox{.}(2021)]%
        {vit}
\bibfield{author}{\bibinfo{person}{Alexey Dosovitskiy}, \bibinfo{person}{Lucas Beyer}, \bibinfo{person}{Alexander Kolesnikov}, \bibinfo{person}{Dirk Weissenborn}, \bibinfo{person}{Xiaohua Zhai}, \bibinfo{person}{Thomas Unterthiner}, \bibinfo{person}{Mostafa Dehghani}, \bibinfo{person}{Matthias Minderer}, \bibinfo{person}{Georg Heigold}, \bibinfo{person}{Sylvain Gelly}, \bibinfo{person}{Jakob Uszkoreit}, {and} \bibinfo{person}{Neil Houlsby}.} \bibinfo{year}{2021}\natexlab{}.
\newblock \showarticletitle{An Image is Worth 16x16 Words: Transformers for Image Recognition at Scale}. In \bibinfo{booktitle}{\emph{International Conference on Learning Representations}}.
\newblock
\urldef\tempurl%
\url{https://openreview.net/forum?id=YicbFdNTTy}
\showURL{%
\tempurl}


\bibitem[Eisenman et~al\mbox{.}(2019)]%
        {assaf:sysml2019}
\bibfield{author}{\bibinfo{person}{Assaf Eisenman}, \bibinfo{person}{Maxim Naumov}, \bibinfo{person}{Darryl Gardner}, \bibinfo{person}{Misha Smelyanskiy}, \bibinfo{person}{Sergey Pupyrev}, \bibinfo{person}{Kim Hazelwood}, \bibinfo{person}{Asaf Cidon}, {and} \bibinfo{person}{Sachin Katti}.} \bibinfo{year}{2019}\natexlab{}.
\newblock \showarticletitle{{Bandana: Using Non-Volatile Memory for Storing Deep Learning Models}}. In \bibinfo{booktitle}{\emph{Proceedings of the Conference on Systems and Machine Learning (SysML'19)}}. \bibinfo{address}{Stanford, CA}.
\newblock


\bibitem[{Examining AMD Radeon Pro SSG: How NAND Changes the GPU Game}({[n.\,d.]})]%
        {ssg-detail}
\bibfield{author}{\bibinfo{person}{{Examining AMD Radeon Pro SSG: How NAND Changes the GPU Game}}.} \bibinfo{year}{[n.\,d.]}\natexlab{}.
\newblock \bibinfo{howpublished}{\url{https://www.tomshardware.com/news/amd-radeon-pro-ssg,32365.html}}.
\newblock


\bibitem[Ganguly et~al\mbox{.}(2019)]%
        {InterplayUVM}
\bibfield{author}{\bibinfo{person}{Debashis Ganguly}, \bibinfo{person}{Ziyu Zhang}, \bibinfo{person}{Jun Yang}, {and} \bibinfo{person}{Rami Melhem}.} \bibinfo{year}{2019}\natexlab{}.
\newblock \showarticletitle{Interplay between Hardware Prefetcher and Page Eviction Policy in CPU-GPU Unified Virtual Memory}. In \bibinfo{booktitle}{\emph{Proceedings of the 46th International Symposium on Computer Architecture}} \emph{(\bibinfo{series}{ISCA '19})}. \bibinfo{publisher}{Association for Computing Machinery}, \bibinfo{address}{Phoenix, Arizona}.
\newblock


\bibitem[{GPUDirect Storage: A Direct Path Between Storage and GPU Memory}({[n.\,d.]})]%
        {gpudirectstorage}
\bibfield{author}{\bibinfo{person}{{GPUDirect Storage: A Direct Path Between Storage and GPU Memory}}.} \bibinfo{year}{[n.\,d.]}\natexlab{}.
\newblock \bibinfo{howpublished}{\url{https://developer.nvidia.com/blog/gpudirect-storage/}}.
\newblock


\bibitem[{Hao} et~al\mbox{.}(2017)]%
        {fpgaaddr:hpca2017}
\bibfield{author}{\bibinfo{person}{Yuchen {Hao}}, \bibinfo{person}{Zhenman {Fang}}, \bibinfo{person}{Glenn {Reinman}}, {and} \bibinfo{person}{Jason {Cong}}.} \bibinfo{year}{2017}\natexlab{}.
\newblock \showarticletitle{Supporting Address Translation for Accelerator-Centric Architectures}. In \bibinfo{booktitle}{\emph{Proceedings of the IEEE International Symposium on High Performance Computer Architecture (HPCA'17)}}.
\newblock


\bibitem[He et~al\mbox{.}(2016)]%
        {ResNet}
\bibfield{author}{\bibinfo{person}{Kaiming He}, \bibinfo{person}{Xiangyu Zhang}, \bibinfo{person}{Shaoqing Ren}, {and} \bibinfo{person}{Jian Sun}.} \bibinfo{year}{2016}\natexlab{}.
\newblock \showarticletitle{{Deep residual learning for image recognition}}. In \bibinfo{booktitle}{\emph{Proceedings of the IEEE Conference on Computer Vision and Pattern Recognition (CVPR'16)}}. \bibinfo{address}{Las Vegas, NV}.
\newblock


\bibitem[{High Bandwidth Memory}({[n.\,d.]})]%
        {hbm}
\bibfield{author}{\bibinfo{person}{{High Bandwidth Memory}}.} \bibinfo{year}{[n.\,d.]}\natexlab{}.
\newblock \bibinfo{howpublished}{\url{https://en.wikipedia.org/wiki/High_Bandwidth_Memory}}.
\newblock


\bibitem[Hildebrand et~al\mbox{.}(2020)]%
        {autotm}
\bibfield{author}{\bibinfo{person}{Mark Hildebrand}, \bibinfo{person}{Jawad Khan}, \bibinfo{person}{Sanjeev Trika}, \bibinfo{person}{Jason Lowe-Power}, {and} \bibinfo{person}{Venkatesh Akella}.} \bibinfo{year}{2020}\natexlab{}.
\newblock \showarticletitle{AutoTM: Automatic Tensor Movement in Heterogeneous Memory Systems Using Integer Linear Programming} \emph{(\bibinfo{series}{ASPLOS '20})}. \bibinfo{publisher}{Association for Computing Machinery}, \bibinfo{address}{New York, NY, USA}, \bibinfo{pages}{875–890}.
\newblock
\showISBNx{9781450371025}
\urldef\tempurl%
\url{https://doi.org/10.1145/3373376.3378465}
\showDOI{\tempurl}


\bibitem[Hu et~al\mbox{.}(2018)]%
        {SENet}
\bibfield{author}{\bibinfo{person}{Jie Hu}, \bibinfo{person}{Li Shen}, {and} \bibinfo{person}{Gang Sun}.} \bibinfo{year}{2018}\natexlab{}.
\newblock \showarticletitle{Squeeze-and-excitation networks}. In \bibinfo{booktitle}{\emph{Proceedings of the IEEE conference on computer vision and pattern recognition}}. \bibinfo{pages}{7132--7141}.
\newblock


\bibitem[Huang et~al\mbox{.}(2020)]%
        {SwapAdvisor}
\bibfield{author}{\bibinfo{person}{Chien-Chin Huang}, \bibinfo{person}{Gu Jin}, {and} \bibinfo{person}{Jinyang Li}.} \bibinfo{year}{2020}\natexlab{}.
\newblock \showarticletitle{SwapAdvisor: Pushing Deep Learning Beyond the GPU Memory Limit via Smart Swapping}. In \bibinfo{booktitle}{\emph{Proceedings of the Twenty-Fifth International Conference on Architectural Support for Programming Languages and Operating Systems}} \emph{(\bibinfo{series}{ASPLOS'20})}. \bibinfo{address}{Lausanne, Switzerland}.
\newblock


\bibitem[Huang et~al\mbox{.}(2015)]%
        {FlashMap}
\bibfield{author}{\bibinfo{person}{Jian Huang}, \bibinfo{person}{Anirudh Badam}, \bibinfo{person}{Moinuddin~K. Qureshi}, {and} \bibinfo{person}{Karsten Schwan}.} \bibinfo{year}{2015}\natexlab{}.
\newblock \showarticletitle{{Unified Address Translation for Memory-mapped SSDs with FlashMap}}. In \bibinfo{booktitle}{\emph{Proceedings of the 42Nd Annual International Symposium on Computer Architecture}} \emph{(\bibinfo{series}{ISCA '15})}. \bibinfo{address}{Portland, OR}, \bibinfo{pages}{580--591}.
\newblock
\urldef\tempurl%
\url{https://doi.org/10.1145/2749469.2750420}
\showDOI{\tempurl}


\bibitem[{Huggingface, 2023. Transformers.}({[n.\,d.]})]%
        {huggingface}
\bibfield{author}{\bibinfo{person}{{Huggingface, 2023. Transformers.}}} \bibinfo{year}{[n.\,d.]}\natexlab{}.
\newblock \bibinfo{howpublished}{\url{https://github.com/huggingface/transformers/tree/main/examples/pytorch}}.
\newblock


\bibitem[Hyun et~al\mbox{.}(2020)]%
        {neummu:asplos2020}
\bibfield{author}{\bibinfo{person}{Bongjoon Hyun}, \bibinfo{person}{Youngeun Kwon}, \bibinfo{person}{Yujeong Choi}, \bibinfo{person}{John Kim}, {and} \bibinfo{person}{Minsoo Rhu}.} \bibinfo{year}{2020}\natexlab{}.
\newblock \showarticletitle{NeuMMU: Architectural Support for Efficient Address Translations in Neural Processing Units}. In \bibinfo{booktitle}{\emph{Proceedings of the Twenty-Fifth International Conference on Architectural Support for Programming Languages and Operating Systems (ASPLOS'20)}}. \bibinfo{address}{Lausanne, Switzerland}.
\newblock


\bibitem[Intel(2018)]%
        {intel:xpoint}
\bibfield{author}{\bibinfo{person}{Intel}.} \bibinfo{year}{2018}\natexlab{}.
\newblock \bibinfo{title}{{3D XPoint: A Breakthrough in Non-Volatile Memory Technology}}.
\newblock \bibinfo{howpublished}{\\\url{https://www.intel.com/content/www/us/en/architecture-and-technology/intel-micron-3d-xpoint-webcast.html}}.
\newblock


\bibitem[Jain et~al\mbox{.}(2020)]%
        {jain:mlsys2020}
\bibfield{author}{\bibinfo{person}{Paras Jain}, \bibinfo{person}{Ajay Jain}, \bibinfo{person}{Aniruddha Nrusimha}, \bibinfo{person}{Amir Gholami}, \bibinfo{person}{Pieter Abbeel}, \bibinfo{person}{Joseph Gonzalez}, \bibinfo{person}{Kurt Keutzer}, {and} \bibinfo{person}{Ion Stoica}.} \bibinfo{year}{2020}\natexlab{}.
\newblock \showarticletitle{Breaking the Memory Wall with Optimal Tensor Rematerialization}.
\newblock In \bibinfo{booktitle}{\emph{Proceedings of Machine Learning and Systems (MLSys'20)}}.
\newblock


\bibitem[Josephson et~al\mbox{.}(2010)]%
        {DFS}
\bibfield{author}{\bibinfo{person}{William~K. Josephson}, \bibinfo{person}{Lars~A. Bongo}, \bibinfo{person}{Kai Li}, {and} \bibinfo{person}{David Flynn}.} \bibinfo{year}{2010}\natexlab{}.
\newblock \showarticletitle{{DFS: A File System for Virtualized Flash Storage}}.
\newblock \bibinfo{journal}{\emph{Trans. Storage}} \bibinfo{volume}{6}, \bibinfo{number}{3}, Article \bibinfo{articleno}{14} (\bibinfo{date}{Sept.} \bibinfo{year}{2010}), \bibinfo{numpages}{25}~pages.
\newblock
\urldef\tempurl%
\url{https://doi.org/10.1145/1837915.1837922}
\showDOI{\tempurl}


\bibitem[Jung et~al\mbox{.}(2023)]%
        {deepum}
\bibfield{author}{\bibinfo{person}{Jaehoon Jung}, \bibinfo{person}{Jinpyo Kim}, {and} \bibinfo{person}{Jaejin Lee}.} \bibinfo{year}{2023}\natexlab{}.
\newblock \showarticletitle{DeepUM: Tensor Migration and Prefetching in Unified Memory}. In \bibinfo{booktitle}{\emph{Proceedings of the 28th ACM International Conference on Architectural Support for Programming Languages and Operating Systems, Volume 2}} (Vancouver, BC, Canada) \emph{(\bibinfo{series}{ASPLOS 2023})}. \bibinfo{publisher}{Association for Computing Machinery}, \bibinfo{address}{New York, NY, USA}, \bibinfo{pages}{207–221}.
\newblock
\showISBNx{9781450399166}
\urldef\tempurl%
\url{https://doi.org/10.1145/3575693.3575736}
\showDOI{\tempurl}


\bibitem[Kim et~al\mbox{.}(2020)]%
        {batchuvm:asplos2020}
\bibfield{author}{\bibinfo{person}{Hyojong Kim}, \bibinfo{person}{Jaewoong Sim}, \bibinfo{person}{Prasun Gera}, \bibinfo{person}{Ramyad Hadidi}, {and} \bibinfo{person}{Hyesoon Kim}.} \bibinfo{year}{2020}\natexlab{}.
\newblock \showarticletitle{Batch-Aware Unified Memory Management in GPUs for Irregular Workloads}. In \bibinfo{booktitle}{\emph{Proceedings of the Twenty-Fifth International Conference on Architectural Support for Programming Languages and Operating Systems (ASPLOS'20)}}. \bibinfo{address}{Lausanne, Switzerland}.
\newblock


\bibitem[Kim et~al\mbox{.}(2014)]%
        {gpunet}
\bibfield{author}{\bibinfo{person}{Sangman Kim}, \bibinfo{person}{Seonggu Huh}, \bibinfo{person}{Xinya Zhang}, \bibinfo{person}{Yige Hu}, \bibinfo{person}{Amir Wated}, \bibinfo{person}{Emmett Witchel}, {and} \bibinfo{person}{Mark Silberstein}.} \bibinfo{year}{2014}\natexlab{}.
\newblock \showarticletitle{{GPUnet: Networking Abstractions for {GPU} Programs}}. In \bibinfo{booktitle}{\emph{Proceedings of the 11th {USENIX} Symposium on Operating Systems Design and Implementation ({OSDI}'14)}}. \bibinfo{address}{Broomfield, CO}.
\newblock


\bibitem[Krizhevsky et~al\mbox{.}(2012)]%
        {krizhevsky2012imagenet}
\bibfield{author}{\bibinfo{person}{Alex Krizhevsky}, \bibinfo{person}{Ilya Sutskever}, {and} \bibinfo{person}{Geoffrey~E Hinton}.} \bibinfo{year}{2012}\natexlab{}.
\newblock \showarticletitle{Imagenet classification with deep convolutional neural networks}. In \bibinfo{booktitle}{\emph{Advances in neural information processing systems}}.
\newblock


\bibitem[Kwon et~al\mbox{.}(2019)]%
        {tensordimm}
\bibfield{author}{\bibinfo{person}{Youngeun Kwon}, \bibinfo{person}{Yunjae Lee}, {and} \bibinfo{person}{Minsoo Rhu}.} \bibinfo{year}{2019}\natexlab{}.
\newblock \showarticletitle{TensorDIMM: A Practical Near-Memory Processing Architecture for Embeddings and Tensor Operations in Deep Learning}. In \bibinfo{booktitle}{\emph{Proceedings of the 52nd Annual IEEE/ACM International Symposium on Microarchitecture (MICRO'19)}}. \bibinfo{address}{Columbus, OH, USA}.
\newblock


\bibitem[Kwon and Rhu(2018)]%
        {memorywall:micro2018}
\bibfield{author}{\bibinfo{person}{Youngeun Kwon} {and} \bibinfo{person}{Minsoo Rhu}.} \bibinfo{year}{2018}\natexlab{}.
\newblock \showarticletitle{Beyond the Memory Wall: A Case for Memory-Centric HPC System for Deep Learning}. In \bibinfo{booktitle}{\emph{Proceedings of the 51st Annual IEEE/ACM International Symposium on Microarchitecture (MICRO'18)}}. \bibinfo{address}{Fukuoka, Japan}.
\newblock


\bibitem[Le et~al\mbox{.}(2019)]%
        {le2019tflms}
\bibfield{author}{\bibinfo{person}{Tung~D. Le}, \bibinfo{person}{Haruki Imai}, \bibinfo{person}{Yasushi Negishi}, {and} \bibinfo{person}{Kiyokuni Kawachiya}.} \bibinfo{year}{2019}\natexlab{}.
\newblock \bibinfo{title}{TFLMS: Large Model Support in TensorFlow by Graph Rewriting}.
\newblock
\newblock
\showeprint[arxiv]{1807.02037}~[cs.LG]


\bibitem[Lew et~al\mbox{.}(2019)]%
        {gpgpusim}
\bibfield{author}{\bibinfo{person}{Jonathan Lew}, \bibinfo{person}{Deval~A. Shah}, \bibinfo{person}{Suchita Pati}, \bibinfo{person}{Shaylin Cattell}, \bibinfo{person}{Mengchi Zhang}, \bibinfo{person}{Amruth Sandhupatla}, \bibinfo{person}{Christopher Ng}, \bibinfo{person}{Negar Goli}, \bibinfo{person}{Matthew~D. Sinclair}, \bibinfo{person}{Timothy~G. Rogers}, {and} \bibinfo{person}{Tor~M. Aamodt}.} \bibinfo{year}{2019}\natexlab{}.
\newblock \showarticletitle{Analyzing Machine Learning Workloads Using a Detailed GPU Simulator}. In \bibinfo{booktitle}{\emph{2019 IEEE International Symposium on Performance Analysis of Systems and Software (ISPASS)}}.
\newblock


\bibitem[Li et~al\mbox{.}(2020)]%
        {ang:gpulink}
\bibfield{author}{\bibinfo{person}{Ang Li}, \bibinfo{person}{Shuaiwen~Leon Song}, \bibinfo{person}{Jieyang Chen}, \bibinfo{person}{Jiajia Li}, \bibinfo{person}{Xu Liu}, \bibinfo{person}{Nathan~R. Tallent}, {and} \bibinfo{person}{Kevin~J. Barker}.} \bibinfo{year}{2020}\natexlab{}.
\newblock \showarticletitle{{Evaluating Modern GPU Interconnect: PCIe, NVLink, NV-SLI, NVSwitch and GPUDirect}}.
\newblock \bibinfo{journal}{\emph{IEEE Transactions on Parallel and Distributed Systems}} \bibinfo{volume}{31}, \bibinfo{number}{1} (\bibinfo{date}{January} \bibinfo{year}{2020}).
\newblock


\bibitem[Loh(2008)]%
        {3dstack:isca2008}
\bibfield{author}{\bibinfo{person}{Gabriel~H. Loh}.} \bibinfo{year}{2008}\natexlab{}.
\newblock \showarticletitle{3D-Stacked Memory Architectures for Multi-Core Processors}. In \bibinfo{booktitle}{\emph{Proceedings of the 35th Annual International Symposium on Computer Architecture (ISCA'08)}}. \bibinfo{address}{USA}.
\newblock


\bibitem[{NVIDIA H100 Tensor Core GPU}({[n.\,d.]})]%
        {h100-hbm}
\bibfield{author}{\bibinfo{person}{{NVIDIA H100 Tensor Core GPU}}.} \bibinfo{year}{[n.\,d.]}\natexlab{}.
\newblock \bibinfo{howpublished}{\url{https://www.nvidia.com/en-us/data-center/h100/}}.
\newblock


\bibitem[Peng et~al\mbox{.}(2018)]%
        {peng2018optimus}
\bibfield{author}{\bibinfo{person}{Yanghua Peng}, \bibinfo{person}{Yixin Bao}, \bibinfo{person}{Yangrui Chen}, \bibinfo{person}{Chuan Wu}, {and} \bibinfo{person}{Chuanxiong Guo}.} \bibinfo{year}{2018}\natexlab{}.
\newblock \showarticletitle{{Optimus: An Efficient Dynamic Resource Scheduler for Deep Learning Clusters}}. In \bibinfo{booktitle}{\emph{Proceedings of the 13th European Conference on Computer Systems (EuroSys'18)}}. \bibinfo{address}{Porto, Portugal}.
\newblock


\bibitem[{PyTorch, 2023. PyTorch Examples}({[n.\,d.]})]%
        {pytorchexamples}
\bibfield{author}{\bibinfo{person}{{PyTorch, 2023. PyTorch Examples}}.} \bibinfo{year}{[n.\,d.]}\natexlab{}.
\newblock \bibinfo{howpublished}{\url{https://pytorch.org/examples/\#pytorch-examples}}.
\newblock


\bibitem[Qureshi et~al\mbox{.}(2022)]%
        {BaM}
\bibfield{author}{\bibinfo{person}{Zaid Qureshi}, \bibinfo{person}{Vikram~Sharma Mailthody}, \bibinfo{person}{Isaac Gelado}, \bibinfo{person}{Seung~Won Min}, \bibinfo{person}{Amna Masood}, \bibinfo{person}{Jeongmin Park}, \bibinfo{person}{Jinjun Xiong}, \bibinfo{person}{CJ Newburn}, \bibinfo{person}{Dmitri Vainbrand}, \bibinfo{person}{I Chung}, {et~al\mbox{.}}} \bibinfo{year}{2022}\natexlab{}.
\newblock \showarticletitle{BaM: A Case for Enabling Fine-grain High Throughput GPU-Orchestrated Access to Storage}.
\newblock \bibinfo{journal}{\emph{arXiv preprint arXiv:2203.04910}} (\bibinfo{year}{2022}).
\newblock


\bibitem[Rajbhandari et~al\mbox{.}(2020)]%
        {ZERO}
\bibfield{author}{\bibinfo{person}{Samyam Rajbhandari}, \bibinfo{person}{Jeff Rasley}, \bibinfo{person}{Olatunji Ruwase}, {and} \bibinfo{person}{Yuxiong He}.} \bibinfo{year}{2020}\natexlab{}.
\newblock \showarticletitle{ZeRO: Memory Optimizations toward Training Trillion Parameter Models}. In \bibinfo{booktitle}{\emph{Proceedings of the International Conference for High Performance Computing, Networking, Storage and Analysis}} (Atlanta, Georgia) \emph{(\bibinfo{series}{SC '20})}. \bibinfo{publisher}{IEEE Press}, Article \bibinfo{articleno}{20}, \bibinfo{numpages}{16}~pages.
\newblock
\showISBNx{9781728199986}


\bibitem[Rajbhandari et~al\mbox{.}(2021)]%
        {zero_infinity}
\bibfield{author}{\bibinfo{person}{Samyam Rajbhandari}, \bibinfo{person}{Olatunji Ruwase}, \bibinfo{person}{Jeff Rasley}, \bibinfo{person}{Shaden Smith}, {and} \bibinfo{person}{Yuxiong He}.} \bibinfo{year}{2021}\natexlab{}.
\newblock \showarticletitle{ZeRO-Infinity: Breaking the GPU Memory Wall for Extreme Scale Deep Learning}. In \bibinfo{booktitle}{\emph{Proceedings of the International Conference for High Performance Computing, Networking, Storage and Analysis}}. \bibinfo{address}{St. Louis, Missouri}.
\newblock


\bibitem[Raoux et~al\mbox{.}(2008)]%
        {mem_pcm_1}
\bibfield{author}{\bibinfo{person}{S. Raoux}, \bibinfo{person}{G.~W. Burr}, \bibinfo{person}{M.~J. Breitwisch}, \bibinfo{person}{C.~T. Rettner}, \bibinfo{person}{Y.~C. Chen}, \bibinfo{person}{R.~M. Shelby}, \bibinfo{person}{M. Salinga}, \bibinfo{person}{D. Krebs}, \bibinfo{person}{S.~H. Chen}, \bibinfo{person}{H.~L. Lung}, {and} \bibinfo{person}{C.~H. Lam}.} \bibinfo{year}{2008}\natexlab{}.
\newblock \showarticletitle{Phase-change random access memory: A scalable technology}.
\newblock \bibinfo{journal}{\emph{IBM Journal of Research and Development}} \bibinfo{volume}{52}, \bibinfo{number}{4.5} (\bibinfo{date}{July} \bibinfo{year}{2008}), \bibinfo{pages}{465--479}.
\newblock
\showISSN{0018-8646}
\urldef\tempurl%
\url{https://doi.org/10.1147/rd.524.0465}
\showDOI{\tempurl}


\bibitem[Ren et~al\mbox{.}(2021a)]%
        {sentinel}
\bibfield{author}{\bibinfo{person}{Jie Ren}, \bibinfo{person}{Jiaolin Luo}, \bibinfo{person}{Kai Wu}, \bibinfo{person}{Minjia Zhang}, \bibinfo{person}{Hyeran Jeon}, {and} \bibinfo{person}{Dong Li}.} \bibinfo{year}{2021}\natexlab{a}.
\newblock \showarticletitle{Sentinel: Efficient Tensor Migration and Allocation on Heterogeneous Memory Systems for Deep Learning}. In \bibinfo{booktitle}{\emph{2021 IEEE International Symposium on High-Performance Computer Architecture (HPCA)}}. \bibinfo{pages}{598--611}.
\newblock
\urldef\tempurl%
\url{https://doi.org/10.1109/HPCA51647.2021.00057}
\showDOI{\tempurl}


\bibitem[Ren et~al\mbox{.}(2021b)]%
        {zero-offload}
\bibfield{author}{\bibinfo{person}{Jie Ren}, \bibinfo{person}{Samyam Rajbhandari}, \bibinfo{person}{Reza~Yazdani Aminabadi}, \bibinfo{person}{Olatunji Ruwase}, \bibinfo{person}{Shuangyan Yang}, \bibinfo{person}{Minjia Zhang}, \bibinfo{person}{Dong Li}, {and} \bibinfo{person}{Yuxiong He}.} \bibinfo{year}{2021}\natexlab{b}.
\newblock \showarticletitle{ZeRO-Offload: Democratizing Billion-Scale Model Training}.
\newblock \bibinfo{journal}{\emph{CoRR}}  \bibinfo{volume}{abs/2101.06840} (\bibinfo{year}{2021}).
\newblock
\urldef\tempurl%
\url{https://arxiv.org/abs/2101.06840}
\showURL{%
\tempurl}


\bibitem[Samsung({[n.\,d.]})]%
        {sumsung-sz985}
\bibfield{author}{\bibinfo{person}{Samsung}.} \bibinfo{year}{[n.\,d.]}\natexlab{}.
\newblock \bibinfo{title}{Samsung Z-SSD SZ985}.
\newblock
\newblock
\urldef\tempurl%
\url{https://semiconductor.samsung.com/resources/brochure/Brochure_Samsung_S-ZZD_SZ985_1804.pdf}
\showURL{%
\tempurl}


\bibitem[{Samsung Z-NAND}({[n.\,d.]})]%
        {znand}
\bibfield{author}{\bibinfo{person}{{Samsung Z-NAND}}.} \bibinfo{year}{[n.\,d.]}\natexlab{}.
\newblock \bibinfo{howpublished}{\url{https://www.samsung.com/semiconductor/ssd/z-ssd/}}.
\newblock


\bibitem[Saxena and Swift(2010)]%
        {FlashVM}
\bibfield{author}{\bibinfo{person}{Mohit Saxena} {and} \bibinfo{person}{Michael~M. Swift}.} \bibinfo{year}{2010}\natexlab{}.
\newblock \showarticletitle{{FlashVM: Virtual Memory Management on Flash}}. In \bibinfo{booktitle}{\emph{Proceedings of the 2010 USENIX Conference on USENIX Annual Technical Conference}} \emph{(\bibinfo{series}{USENIXATC'10})}. \bibinfo{address}{Boston, MA}, \bibinfo{pages}{187--200}.
\newblock


\bibitem[Shahar et~al\mbox{.}(2016)]%
        {activepointer:isca2016}
\bibfield{author}{\bibinfo{person}{Sagi Shahar}, \bibinfo{person}{Shai Bergman}, {and} \bibinfo{person}{Mark Silberstein}.} \bibinfo{year}{2016}\natexlab{}.
\newblock \showarticletitle{ActivePointers: A Case for Software Address Translation on GPUs}. In \bibinfo{booktitle}{\emph{Proceedings of the 43rd International Symposium on Computer Architecture (ISCA'16)}}. \bibinfo{address}{Seoul, Republic of Korea}.
\newblock


\bibitem[Silberstein et~al\mbox{.}(2013)]%
        {gpufs}
\bibfield{author}{\bibinfo{person}{Mark Silberstein}, \bibinfo{person}{Bryan Ford}, \bibinfo{person}{Idit Keidar}, {and} \bibinfo{person}{Emmett Witchel}.} \bibinfo{year}{2013}\natexlab{}.
\newblock \showarticletitle{{GPUfs: integrating file systems with GPUs}}. In \bibinfo{booktitle}{\emph{Proceedings of the 18th International Conference on Architectural Support for Programming Languages and Operating Systems (ASPLOS'13)}}. \bibinfo{address}{Houston, Texas, USA}.
\newblock


\bibitem[Szegedy et~al\mbox{.}(2017)]%
        {szegedy2017inception}
\bibfield{author}{\bibinfo{person}{Christian Szegedy}, \bibinfo{person}{Sergey Ioffe}, \bibinfo{person}{Vincent Vanhoucke}, {and} \bibinfo{person}{Alexander~A Alemi}.} \bibinfo{year}{2017}\natexlab{}.
\newblock \showarticletitle{Inception-v4, inception-resnet and the impact of residual connections on learning}. In \bibinfo{booktitle}{\emph{Thirty-first AAAI conference on artificial intelligence}}.
\newblock


\bibitem[Szegedy et~al\mbox{.}(2016)]%
        {szegedy2016rethinking}
\bibfield{author}{\bibinfo{person}{Christian Szegedy}, \bibinfo{person}{Vincent Vanhoucke}, \bibinfo{person}{Sergey Ioffe}, \bibinfo{person}{Jon Shlens}, {and} \bibinfo{person}{Zbigniew Wojna}.} \bibinfo{year}{2016}\natexlab{}.
\newblock \showarticletitle{Rethinking the inception architecture for computer vision}. In \bibinfo{booktitle}{\emph{Proceedings of the IEEE conference on computer vision and pattern recognition}}. \bibinfo{pages}{2818--2826}.
\newblock


\bibitem[Tan and Le(2019)]%
        {tan2019efficientnet}
\bibfield{author}{\bibinfo{person}{Mingxing Tan} {and} \bibinfo{person}{Quoc Le}.} \bibinfo{year}{2019}\natexlab{}.
\newblock \showarticletitle{Efficientnet: Rethinking model scaling for convolutional neural networks}. In \bibinfo{booktitle}{\emph{International conference on machine learning}}. PMLR, \bibinfo{pages}{6105--6114}.
\newblock


\bibitem[{The World's First GPU to Break the Terabyte Memory Barrier}({[n.\,d.]})]%
        {ssg}
\bibfield{author}{\bibinfo{person}{{The World's First GPU to Break the Terabyte Memory Barrier}}.} \bibinfo{year}{[n.\,d.]}\natexlab{}.
\newblock \bibinfo{howpublished}{\url{https://www.amd.com/en/products/professional-graphics/radeon-pro-ssg}}.
\newblock


\bibitem[{Unified CPU/GPU Memory Architecture Raises the Performance Bar}({[n.\,d.]})]%
        {amduvm}
\bibfield{author}{\bibinfo{person}{{Unified CPU/GPU Memory Architecture Raises the Performance Bar}}.} \bibinfo{year}{[n.\,d.]}\natexlab{}.
\newblock \bibinfo{howpublished}{\url{https://www.electronicdesign.com/technologies/microcontrollers/article/21796296/unified-cpugpu-memory-architecture-raises-the-performance-bar}}.
\newblock


\bibitem[{Unified Memory for CUDA Beginners}({[n.\,d.]})]%
        {uvm}
\bibfield{author}{\bibinfo{person}{{Unified Memory for CUDA Beginners}}.} \bibinfo{year}{[n.\,d.]}\natexlab{}.
\newblock \bibinfo{howpublished}{\url{https://developer.nvidia.com/blog/unified-memory-cuda-beginners/}}.
\newblock


\bibitem[Warstadt et~al\mbox{.}(2018)]%
        {warstadt2018neural}
\bibfield{author}{\bibinfo{person}{Alex Warstadt}, \bibinfo{person}{Amanpreet Singh}, {and} \bibinfo{person}{Samuel~R Bowman}.} \bibinfo{year}{2018}\natexlab{}.
\newblock \showarticletitle{Neural Network Acceptability Judgments}.
\newblock \bibinfo{journal}{\emph{arXiv preprint arXiv:1805.12471}} (\bibinfo{year}{2018}).
\newblock


\bibitem[Xie et~al\mbox{.}(2017)]%
        {ResNeXt}
\bibfield{author}{\bibinfo{person}{Saining Xie}, \bibinfo{person}{Ross Girshick}, \bibinfo{person}{Piotr Doll{\'a}r}, \bibinfo{person}{Zhuowen Tu}, {and} \bibinfo{person}{Kaiming He}.} \bibinfo{year}{2017}\natexlab{}.
\newblock \showarticletitle{Aggregated residual transformations for deep neural networks}. In \bibinfo{booktitle}{\emph{Proceedings of the IEEE conference on computer vision and pattern recognition}}. \bibinfo{pages}{1492--1500}.
\newblock


\bibitem[Zagoruyko and Komodakis(2016)]%
        {zagoruyko2016wide}
\bibfield{author}{\bibinfo{person}{Sergey Zagoruyko} {and} \bibinfo{person}{Nikos Komodakis}.} \bibinfo{year}{2016}\natexlab{}.
\newblock \showarticletitle{Wide residual networks}.
\newblock \bibinfo{journal}{\emph{arXiv preprint arXiv:1605.07146}} (\bibinfo{year}{2016}).
\newblock


\bibitem[{Zhang} and {Jung}(2020)]%
        {zng:isca2020}
\bibfield{author}{\bibinfo{person}{Jie {Zhang}} {and} \bibinfo{person}{Myoungsoo {Jung}}.} \bibinfo{year}{2020}\natexlab{}.
\newblock \showarticletitle{ZnG: Architecting GPU Multi-Processors with New Flash for Scalable Data Analysis}. In \bibinfo{booktitle}{\emph{Proceedings of the ACM/IEEE 47th Annual International Symposium on Computer Architecture (ISCA'20)}}.
\newblock


\bibitem[Zhang et~al\mbox{.}(2019)]%
        {flashgpu}
\bibfield{author}{\bibinfo{person}{Jie Zhang}, \bibinfo{person}{Miryeong Kwon}, \bibinfo{person}{Hyojong Kim}, \bibinfo{person}{Hyesoon Kim}, {and} \bibinfo{person}{Myoungsoo Jung}.} \bibinfo{year}{2019}\natexlab{}.
\newblock \showarticletitle{FlashGPU: Placing New Flash Next to GPU Cores}. In \bibinfo{booktitle}{\emph{Proceedings of the 56th Annual Design Automation Conference (DAC'19)}} (Las Vegas, NV, USA).
\newblock


\bibitem[Zhang et~al\mbox{.}(2012)]%
        {namelesswrite:fast2012}
\bibfield{author}{\bibinfo{person}{Yiying Zhang}, \bibinfo{person}{Leo~Prasath Arulraj}, \bibinfo{person}{Andrea~C. Arpaci-Dusseau}, {and} \bibinfo{person}{Remzi~H. Arpaci-Dusseau}.} \bibinfo{year}{2012}\natexlab{}.
\newblock \showarticletitle{{De-indirection for Flash-based SSDs with Nameless Writes}}. In \bibinfo{booktitle}{\emph{Proc. 10th {USENIX} {FAST}}}. \bibinfo{address}{San Jose, {CA}}.
\newblock


\bibitem[Zhao et~al\mbox{.}(2013)]%
        {jishen:cgo}
\bibfield{author}{\bibinfo{person}{Jishen Zhao}, \bibinfo{person}{Guangyu Sun}, \bibinfo{person}{Gabriel~H. Loh}, {and} \bibinfo{person}{Yuan Xie}.} \bibinfo{year}{2013}\natexlab{}.
\newblock \showarticletitle{Optimizing GPU Energy Efficiency with 3D Die-Stacking Graphics Memory and Reconfigurable Memory Interface}.
\newblock \bibinfo{journal}{\emph{ACM Trans. Archit. Code Optim.}} \bibinfo{volume}{10}, \bibinfo{number}{4}, Article \bibinfo{articleno}{24} (\bibinfo{date}{Dec.} \bibinfo{year}{2013}).
\newblock


\bibitem[Zheng et~al\mbox{.}(2016)]%
        {towardsUVM}
\bibfield{author}{\bibinfo{person}{Tianhao Zheng}, \bibinfo{person}{David Nellans}, \bibinfo{person}{Arslan Zulfiqar}, \bibinfo{person}{Mark Stephenson}, {and} \bibinfo{person}{Stephen~W. Keckler}.} \bibinfo{year}{2016}\natexlab{}.
\newblock \showarticletitle{Towards high performance paged memory for GPUs}. In \bibinfo{booktitle}{\emph{2016 IEEE International Symposium on High Performance Computer Architecture (HPCA)}}. \bibinfo{pages}{345--357}.
\newblock
\urldef\tempurl%
\url{https://doi.org/10.1109/HPCA.2016.7446077}
\showDOI{\tempurl}


\end{thebibliography}
\flushcolsend

\newpage

\appendix

\lstdefinestyle{BashStyle} {frame=tb,
  language=bash,
  aboveskip=3mm,
  belowskip=3mm,
  showstringspaces=false,
  columns=flexible,
  basicstyle={\scriptsize\ttfamily},
  numbers=left,
  numbersep=5pt,
  numberstyle=\tiny\color{gray},
  keywordstyle=\color{blue},
  commentstyle=\color{commentgreen},
  stringstyle=\color{mauve},
  breaklines=true,
  breakatwhitespace=true,
  tabsize=4,
  classoffset=0,
  morekeywords={apt, pip3, make, cd, wget},
  keywordstyle=\color{blue},
}

\section{Artifact Appendix}

\subsection{Abstract}

We implement \pname{} by building our own simulation framework described in ($\S$\ref{sec:implementation}). In this artifact, we provide the source code of \pname{} and necessary instructions to reproduce key performance results (Figure 2-4 in $\S$\ref{sec:study} and Figure 11-19 in $\S$\ref{sec:eval}). 

The artifact can be executed on any x86 machine with at least 30 GB of main memory and at least 120 GB of disk space. We strongly recommend running the artifact on a workstation with multi-cores and at least 128 GB memory.

\subsection{Artifact Checklist (Meta-Information)}

{\small
\begin{itemize}[leftmargin=*]
  \item {\bf Algorithm:} Tensor Vitality Analysis and Smart Tensor Migration Scheduling Algorithm.
  \item {\bf Compilation:} GCC 9.4.0 or newer versions.
  \item {\bf Neural Network Models:} BERT, ViT, ResNet, Inceptionv3, SENet. Their traces are included in the repo.
  \item {\bf Run-time environment:} Ubuntu 18.04 or newer versions.
  \item {\bf Metrics:} Execution time, training throughput, and migration traffic.
  \item {\bf Output:} Files and graphs, expected results included in the repo.
  \item {\bf Experiments:} Generate experiments using supplied scripts.
  \item {\bf How much disk space required (approximately):} 120 GB
  \item {\bf How much time is needed to prepare workflow (approximately):} 10 mins
  \item {\bf How much time is needed to complete experiments (approximately): }
  20 hours on a server with 256 GB main memory.
  \item {\bf Publicly available:} Yes
  \item {\bf Archived (provide DOI):} 10.5281/zenodo.8294395
\end{itemize}
}

\subsection{Description}

\subsubsection{How to Access}

The source code can be downloaded from Zenodo at \url{https://doi.org/10.5281/zenodo.8294395}. For the latest version, you can access our GitHub repository: \url{https://github.com/platformxlab/G10.git}.

\subsubsection{Hardware Dependencies}

The artifact can be executed on any x86 machine with at least 30 GB of main memory and at least 120 GB of disk space.

\subsubsection{Software Dependencies}

The artifact needs a Linux environment (preferably Ubuntu) with C++ 14 standard compilation supported.

\subsection{Installation}

\begin{enumerate}
    \item Start by downloading the \pname{} artifact from Zenodo:
    \begin{lstlisting}[style=BashStyle,escapechar=~~,label=code:instrumentation]
    wget https://zenodo.org/record/8294395/files/G10-Artifact.tar.gz
    tar -xvf G10-Artifact.tar.gz
    \end{lstlisting}
    \item Please make sure all prerequisites are successfully installed:
    \begin{lstlisting}[style=BashStyle,escapechar=~~,label=code:instrumentation]
    sudo apt install flex bison tmux python3-pip
    pip3 install matplotlib networkx pandas PyPDF2
    \end{lstlisting}
    \item Build \pname{} (the output executable is named \textit{gpg}):
    \begin{lstlisting}[style=BashStyle,escapechar=~~,label=code:instrumentation]
    cd G10-Artifact/src
    make clean && make
    \end{lstlisting}
\end{enumerate}

\subsection{Experiment Workflow}

This section describes the steps to generate and run the necessary experiments. We strongly recommend readers to follow \textit{"resources/README.md"} to understand more about each script used in this section.

\subsubsection{Generating Configurations} 
The first step is to generate appropriate config files. In this artifact, we provide the Python script \textit{"resources/genconfigs.py"} to generate all the config files used in this artifact (in \textit{configs/} directory). 
\begin{lstlisting}[style=BashStyle,escapechar=~~,label=code:instrumentation]
    python3 resources/genconfigs.py
\end{lstlisting}

\subsubsection{Launching A Single Experiment} 
Every configuration file specifies the DNN model and the batch size to be used, as well as other system configuration parameters (such as GPU memory size, SSD Bandwidth, the baseline type, and so on). All the DNN model graph information and their execution traces are already included if users use the configs generated by the \textit{"resources/genconfigs.py"} script.

To run a single experiment, directly find its corresponding config file and use ((use G10-(BERT, batchsize=256)) as an example):
\begin{lstlisting}[style=BashStyle,escapechar=~~,label=code:instrumentation]
    ./gpg "$relative_path_to_config_file"
    # e.g.,  ./gpg configs/BERT/256-sim_prefetch_lru.config 
\end{lstlisting}

The program will execute the Tensor Vitality Analysis and Smart Tensor Migration Algorithms, and do a performance simulation of the DNN training. The results will be placed 
under the \textit{G10-Artifact/results} directory. 

For each experiment, G10 will generate separate logs for analyzed DNN graph information, tensor vitality analysis results, smart tensor migration scheduling, and performance simulation results.

See \textit{"G10-Artifact/results/README.md"} for more details of the experiment output.

\subsubsection{Launching Batched Experiments}
To run a large number of experiments at one time, we provide the \textit{"resources/run.sh"} Shell script. It can use regular expressions to match multiple config files, and it will automatically spawn different experiments to multiple \textit{tmux} windows for parallel execution.

To evaluate all the experiments more conveniently, we provide a Shell script, \textit{"artifact\_run.sh"}, which will be introduced in the next section. To run individual experiments corresponding to the figures in the paper, see lines 23-45 of \textit{"artifact\_run.sh"}:
\begin{lstlisting}[style=BashStyle,escapechar=~~,label=code:instrumentation]
    # First run experiments for figure 11-14
    ./run.sh -p "(BERT\/256|VIT\/1280|Inceptionv3\/1536|
        ResNet152\/1280|SENet154\/1024)-sim_
        (deepUM|prefetch_lru|FlashNeuron|G10GDSSSD|G10GDSFULL|lru)
        \.config" 
        -dr -j $MAX_PROCESS_NUM
    
    # Then run experiments for figure 15
    ./run.sh -p "(BERT\/(128|256|512|768|1024)|
        VIT\/(256|512|768|1024|1280)|
        Inceptionv3\/(512|768|1024|1280|1536|1792)|
        ResNet152\/(256|512|768|1024|1280)|
        SENet154\/(256|512|768|1024))
        -sim_(deepUM|prefetch_lru|FlashNeuron|lru)\.config" 
        -dr -j $MAX_PROCESS_NUM
    
    # Then run experiments for figure 16
    ./run.sh -p "(BERT\/(256|384|512|640)|
        VIT\/(768|1024|1280|1536)|
        Inceptionv3\/(512|1024|1280|1536)|
        ResNet152\/(768|1024|1280|1536)|
        SENet154\/(256|512|768|1024))
        -sim_prefetch_lru
        (-cpu(0|16|32|64|96|192|256))?\.config" 
        -dr -j $MAX_PROCESS_NUM
    
    # Then run experiments for figure 17
    ./run.sh -p "(VIT\/1024|Inceptionv3\/1280)
        -sim_(deepUM|prefetch_lru|FlashNeuron)
        -cpu(0|16|32|64|256)\.config" 
        -dr -j $MAX_PROCESS_NUM
    
    # Then run experiments for figure 18
    ./run.sh -p "(BERT\/512|VIT\/1280|Inceptionv3\/1536|
        ResNet152\/1280|SENet154\/1024)
        -sim_(deepUM|prefetch_lru|FlashNeuron|lru)
        -ssd(6_4|12_8|19_2|25_6|32)-.*\.config" 
        -dr -j $MAX_PROCESS_NUM
    
    # Then run experiments for figure 19
    ./run.sh -p "(BERT\/256|VIT\/1280|Inceptionv3\/1536|
        ResNet152\/1280|SENet154\/1024)
        -sim_prefetch_lru-var0_(05|10|15|20|25)\.config" 
        -dr -j $MAX_PROCESS_NUM

\end{lstlisting}

\subsection{Evaluation and Expected Results}

To evaluate the artifact results, simply run:
\begin{lstlisting}[style=BashStyle,escapechar=~~,label=code:instrumentation]
    ./artifact_run.sh
\end{lstlisting}

This script runs all the experiments, data gathering, and figure drawing sequentially. Note that users may have to change the maximum allowed number of 
parallel experiments (i.e., the variable \textit{\$MAX\_PROCESS\_NUM}) in the script, based on the machine's main memory capacity (each process 
needs a peak memory of about 28.5 GB). A detailed description of each command and the location of the output figures are also included in the script.

We have provided the expected result files in the directory \textit{"G10-Artifact/example\_results"}. To verify the results, one can compare the 
generated figures directly with those in the paper, or compare the data for each figure with the example results we provided.

\newpage
\subsection{Experiment Customization}

\subsubsection{Changing Simulation Configurations}
In addition to the provided configurations, users can also customize their own config files and evaluate them. The simplest way to do this is to modify the \textit{"resources/genconfigs.py"} script. Note that we only provided DNN training execution traces for some specific batch sizes.

\subsubsection{Custom DNN Training Profiling}
Users can generate their own traces of DNN training on their own GPUs. Users can also generate traces for customized batch sizes. 
Custom profiling can be done by modifying the config files named "profile" rather than "sim", and running them with the \pname{} executable. 
Note that to do this, users have to first correctly install \textit{CUDA} (11.0 and newer version) tool-kits with \textit{cudnn} and \textit{cublas} libraries. 
Before the custom profiling, please make sure you have built the CUDA code generation part of our framework:
\begin{lstlisting}[style=BashStyle,escapechar=~~,label=code:instrumentation]
    cd G10-Artifact/src/cudnn
    make clean && make
\end{lstlisting}
Note that the profiling may take a long time.

\subsection{Methodology}

Submission, reviewing and badging methodology:

\begin{itemize}
  \item \url{https://www.acm.org/publications/policies/artifact-review-and-badging-current}
  \item \url{http://cTuning.org/ae/submission-20201122.html}
  \item \url{http://cTuning.org/ae/reviewing-20201122.html}
\end{itemize}

\end{document}